\documentclass[sigconf,nonacm]{acmart}
\usepackage{subcaption}
\usepackage{lscape}
\usepackage{enumitem}
\usepackage{tabularx}
\usepackage{multirow}

\AtBeginDocument{%
  }

\copyrightyear{2026}
\acmYear{2026}
\setcopyright{cc}
\setcctype{by}
\acmConference[IUI '26]{31st International Conference on Intelligent User Interfaces}{March 23--26, 2026}{Paphos, Cyprus}
\acmBooktitle{31st International Conference on Intelligent User Interfaces (IUI '26), March 23--26, 2026, Paphos, Cyprus}
\acmPrice{}
\acmDOI{10.1145/3742413.3789123}
\acmISBN{979-8-4007-1984-4/2026/03}

\begin{document}

\title{PersonaMail: Learning and Adapting Personal Communication Preferences for Context-Aware Email Writing}

\author{Rui Yao}
\affiliation{%
  \institution{City University of Hong Kong}
  \city{Hong Kong}
  \country{China}}
\email{yaorui2-c@my.cityu.edu.hk}

\author{Qiuyuan Ren}
\affiliation{%
  \institution{City University of Hong Kong}
  \city{Hong Kong}
  \country{China}
}
\email{qiuyuanren1001@gmail.com}

\author{Felicia Fang-Yi Tan}
\affiliation{%
  \institution{New York University}
  \city{New York}
  \country{USA}
  }
  \email{felicia.tan@nyu.edu}
  
\author{Chen Yang}
\affiliation{%
 \institution{City University of Hong Kong}
 \city{Hong Kong}
 \country{China}
 }
\email{sonnechen95@gmail.com}

\author{Xiaoyu Zhang}
\affiliation{%
  \institution{City University of Hong Kong}
  \city{Hong Kong}
  \country{China}}
\email{xiaoyu.zhang@cityu.edu.hk}

\author{Shengdong Zhao}
\authornote{Corresponding author.}
\affiliation{%
  \institution{City University of Hong Kong}
  \city{Hong Kong}
  \country{China}}
\email{shengdong.zhao@cityu.edu.hk}


\begin{abstract}
LLM-assisted writing has seen rapid adoption in interpersonal communication, yet current systems often fail to capture the subtle tones essential for effectiveness. Email writing exemplifies this challenge: effective messages require careful alignment with intent, relationship, and context beyond mere fluency. Through formative studies, we identified three key challenges: articulating nuanced communicative intent, making modifications at multiple levels of granularity, and reusing effective tone strategies across messages. We developed PersonaMail, a system that addresses these gaps through structured communication factor exploration, granular editing controls, and adaptive reuse of successful strategies. Our evaluation compared PersonaMail against standard LLM interfaces, and showed improved efficiency in both immediate and repeated use, alongside higher user satisfaction. We contribute design implications for AI-assisted communication systems that prioritize interpersonal nuance over generic text generation.
\end{abstract}
\begin{teaserfigure}
  \includegraphics[width=\textwidth]{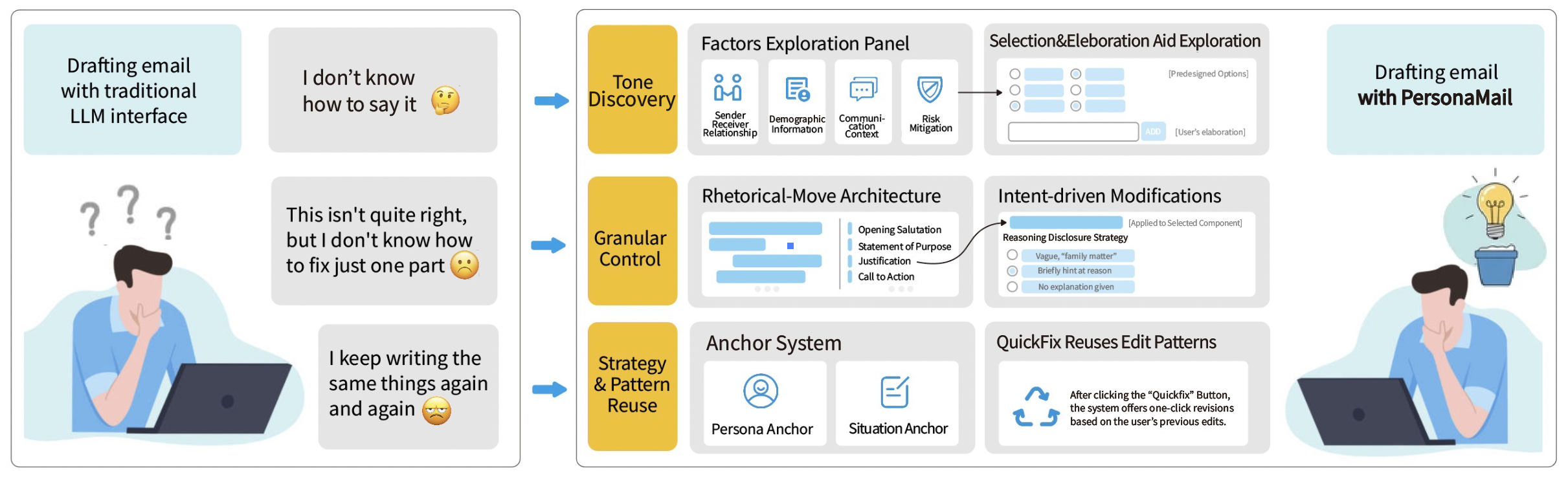}
  \caption{The PersonaMail system overview, contrasting the challenges of drafting emails with traditional LLM interfaces (left) with the structured solutions provided by PersonaMail (center). The system addresses user needs for tone discovery, granular control, and strategy reuse, leading to a more confident and effective writing process.}
    \Description{A three-panel illustration showing the evolution from traditional LLM-assisted email drafting to PersonaMail. Left panel shows a person looking confused with question marks, accompanied by speech bubbles expressing uncertainty about phrasing and repetitive writing. Center panel displays three golden cards labeled "Tone Discovery", "Granular Control", and "Strategy & Pattern Reuse", each connected to detailed interface components: Factors Exploration Panel, Rhetorical-Move Architecture, and Anchor System with their respective features. Right panel shows the same person now confidently working with a lightbulb icon above, representing improved email drafting with PersonaMail.}
  \label{fig:teaser}
\end{teaserfigure}


\keywords{AI-assisted writing, email composition, tone customization, large language models, human-AI collaboration, interpersonal communication, adaptive interfaces}

\maketitle
\footnotetext{This is the author's version of the work. It is posted here for your personal use. Not for redistribution. The definitive Version of Record will be published in ACM IUI 2026.}
\section{Introduction}
Large Language Models (LLMs) have transformed writing across domains, demonstrating remarkable fluency in generating coherent text \cite{kaur2024text}. Their adoption spans creative platforms, business tools, and educational applications, fundamentally changing how we approach text generation \cite{hadi2023large}.

While increasingly useful, LLMs face a persistent challenge: they excel at producing generic, population-level text patterns but are less effective with the personalized nuance that defines meaningful human communication \cite{liu2025survey}. This tension becomes particularly pronounced in high-stakes communication\cite{farrell2015difficult}—managing important personal relationships, navigating difficult conversations, or addressing situations that require careful personal consideration and attention to social nuance.

Consider Dr. Sarah Chen, a postdoc who must withdraw from a conference presentation she fought hard to secure six months ago. Her withdrawal email requires delicate navigation: acknowledging disappointment and inconvenience, justifying her decision without oversharing, maintaining the collegial relationship despite breaking a commitment, all in her established communication style. The stakes feel high—this conference could have advanced her career, and the organizer's opinion matters for future opportunities.

This scenario highlights a fundamental challenge in AI-assisted communication: how do we design systems that help users communicate effectively while honoring their personal preferences, communication style, and situational context? To systematically understand this problem and inform the design of better support tools, we aim to investigate three research questions:

\textbf{RQ1:} What cognitive challenges do users face when writing tone-sensitive emails with LLMs?

\textbf{RQ2:} What factors shape interpersonal tone and communication strategy?

\textbf{RQ3:} How can these insights be translated into the design of AI-assisted writing systems?

Through a formative study, we find current LLMs miss three critical dimensions. First, users often don't know exactly what they want—communicative goals emerge through iteration rather than upfront specification. Users may recognize a message doesn't feel right but struggle to articulate why. Second, existing systems provide insufficient support for revision at different granularities—users need both global changes (adjust formality) and local ones (modify this paragraph's tone). Third, insights like greeting styles or relationship-specific phrasing could transfer to future communications, but existing LLMs don't facilitate reuse.

To address these gaps, we developed PersonaMail with three key innovations. A Factors Exploration Panel supports tone discovery by considering factors through our curated factor list derived from communication research, such as relationships, demographics, context, and risk mitigation. A Communicative-Unit Architecture provides granular control by breaking emails into components (opening, purpose, justification), enabling targeted modifications through manipulation of identified communication intents. Finally, Anchors and QuickFix features enable strategy reuse, allowing users to save strategies for different personas and situations, and apply one-click revisions based on previous editing patterns.

Our evaluation with 16 participants demonstrates PersonaMail's effectiveness. Structured factor exploration produced drafts rated 34.8\% higher in quality than standard LLM interfaces (M=5.81/7 vs. 4.31/7, p<0.001) while reducing cognitive load by 24.5\%. Intent-driven modifications and QuickFix enabled 61.94\% of revisions through targeted adjustments, halving revision time. Adaptive reuse reduced setup time by 58.5\% in subsequent uses, with 42\% faster task completion. These results demonstrate that making implicit communicative knowledge explicit enables more effective human-AI collaboration in writing tasks.

The primary contributions of this work are threefold:

\begin{enumerate}
  \item The identification and characterization of key challenges in AI-assisted tone-sensitive writing through formative studies
  \item The design and development of PersonaMail, a functional prototype that addresses these challenges through novel interaction mechanisms
  \item The evaluation of our system through comparative user studies
\end{enumerate}

\section{Related Work}
\subsection{Tone in Computer-Mediated Communication and Difficult Conversations}

Conveying appropriate tone is integral to clarity, emotional connection, and effective interpersonal outcomes in computer-mediated communication\cite{yuan2020put}. When tone is misinterpreted, feedback loops break down and conflict or confusion can result. Email incivility—often a tone issue—is widespread: over 90\% of professionals report receiving disrespectful-sounding emails, partly because the absence of tonal cues makes neutral messages seem rude \cite{yuan2020put}.

Computer-mediated channels present inherent difficulties for tone expression. Interpersonal communication theories emphasize that written tone reflects complex social cues that are particularly strained in digital media. Walther's hyperpersonal model observes that, lacking nonverbal signals, email senders must strategically craft language through editing time, word choice, and sentence complexity to manage impressions and express relational intent \cite{walther2007selective}. Similarly, Kruger et al. found that people often overestimate their ability to express tone in email because they "hear" their own meaning internally, while recipients lack those interpretive cues \cite{kruger2005egocentrism}.

In difficult conversations, these challenges become even more pronounced. Kippist and Duarte define workplace difficult conversations as those that "potentially make people feel uncomfortable, generate conflict or detrimentally affect relationships" \cite{kippist2015does}. Research across workplace, educational, and personal domains shows that difficult conversations—such as delivering bad news, providing critical feedback, or discussing mistakes and conflicts—are inherently emotionally charged and cognitively taxing. Such exchanges evoke anxiety and risk harming relationships.  Consequently, most critical work occurs even before the first word is spoken, as thorough preparation shifts focus from reactive, emotional confrontation to proactive, purposeful dialogue \cite{soehner2016effective}. This cognitive burden motivates our exploration of how AI-assisted writing tools can better support users in navigating the complexities of tone-sensitive communication.

\subsection {Homogenization Effects in LLM-Mediated Communication}

Large language models demonstrate significant power in simulating human communication and are increasingly involved in people's daily interpersonal interactions as communication agents \cite{fu2024text}. However, a growing body of work finds that AI writing assistance tends to flatten linguistic diversity, diluting individual tone and style. Controlled studies reveal that when multiple users rely on the same model, their texts become more similar. In one experiment, users wrote argumentative essays with or without model assistance; co-writing with an instruction-tuned LLM (InstructGPT) significantly reduced lexical and conceptual diversity and made different authors' essays more alike \cite{padmakumar2023does}. The model tends to produce average phrasing, smoothing out authors' unique stylistic markers. Recent linguistic analyses also confirm that even state-of-the-art LLMs do not fully capture human style variation: systematic differences remain between grammatical and rhetorical features of AI-generated versus human-written text, suggesting LLM output occupies a narrower style space \cite{reinhart2025llms}.

Collectively, these findings indicate that AI suggestions homogenize writing, effectively making diverse voices sound more alike. This homogenization undermines personalization: an LLM may compose a polite, generic email while erasing the personal inflections—casualness, humor, cultural references—that a given writer would naturally use. This limitation makes it particularly difficult to effectively convey nuanced interpersonal communication, especially in difficult conversations where subtle tonal choices can determine communicative success or failure.

\subsection{User Interfaces for LLM-Supported Email Composition}

 Controlling tone in AI-assisted writing remains a significant challenge for both research prototypes and commercial applications. Current approaches predominantly rely on explicit tone selection mechanisms, where users choose from predefined tone categories rather than articulating their unique context-specific preferences.

\subsubsection{Tone Control Mechanisms}
Current approaches to tone control predominantly rely on predefined categories, where users select from fixed catalogs of tone options like "professional," "friendly," or "urgent." This approach is exemplified by commercial tools such as MailMaestro\cite{nMailMaestro}, FridayEmail \cite{FridayEmail}, and Wordtune\cite{Wordtune}, as well as specialized systems like Lampost\cite{goodman_lampost_2024}, which enables localized tone adjustments for adults with dyslexia. Several variants provide additional flexibility: binary toggle mechanisms (Lu et al.\cite{lu_corporate_2024} implement four toggles creating 16 combinations; Masson et al.\cite{masson_textoshop_2025} use continuous scales between paired opposites), and figure-based anchoring where WriteMail \cite{WriteMail} associates styles with public figures like "Anthony Bourdain (Descriptive)" or "Elon Musk (Innovative)." However, all these approaches depends on discrete, predefined categories that may not capture the contextual nuances of interpersonal communication, where tone depends heavily on relationship dynamics, cultural context, and specific situational factors that resist categorization\cite{larkey1996toward}.

\subsubsection{Revision Techniques}
Beyond tone, revision techniques are critical for refining AI-generated text to align with user intent. Most contemporary tools move beyond simple manual editing by incorporating AI-powered local revision functions. Commercial applications like MailMaestro, Friday Email, and WordTune offer a suite of such features, including AI-based proofreading, polishing, and functions like "Extend," "Rewrite," and "Shorten" applied to user-selected text. WordTune further enhances this with suggestions for synonyms and alternative phrasing ("Variants"). Research prototypes explore more sophisticated paradigms. CDLR \cite{zindulka_content-driven_2025}, for instance, allows for prompt-based refinement, enabling a conversational approach to revision , while Textoshop introduces creative techniques inspired by drawing software to give users more expressive control\cite{masson_textoshop_2025}.

\subsubsection{Adaptation and Reuse}
A third dimension of user support is adaptation and the reuse of effective strategies across communications. Our review reveals that this capability is the most underdeveloped in the current design space. The majority of systems offer no mechanism for saving and reapplying successful communication approaches. The primary exception is MailMaestro, which supports variable-based templates and text shortcuts for recurring content like name or addresses. While useful for boilerplate text, this approach is fundamentally limited: it preserves content structure but fails to capture the underlying communicative strategy—the nuanced tone, phrasing, and framing that made a prior email effective. Consequently, users are forced to articulate similar preferences and rediscover effective tone strategies for each new communication, even in recurring scenarios, creating significant and unnecessary cognitive overhead.

Table \ref{tab:designspace} (Appendix B) presents a classification of reviewed email systems based on their tone control, revision, and adaptation capabilities. Although Textoshop is primarily designed for general writing goals, we include it in this table because of its distinctive and representative tone-setting strategy.

\section{Formative Study}
\label{formativestudy}
To understand the cognitive challenges users face when writing tone-sensitive emails with LLMs, we conducted a formative study. After mapping the existing approaches outlined in Table \ref{tab:designspace} (Appendix B), we selected three representative systems - MailMaestro, Friday Email, and Textoshop (Figure~\ref{fig:formativestudy} in Appendix A) - to capture diversity across three critical dimensions: tone control mechanisms, revision techniques, and adaptation capabilities. MailMaestro and Friday Email exemplify commercial tone-selection tools with variable template and categorical tone features; Textoshop represents an experimental, fine-grained tone control paradigm using continuous blending.

In addition, we included Google Gemini paired with Gmail as a baseline workflow, representing the most common real-world practice of AI-assisted email composition through conversational prompting and manual editing.

Detailed descriptions of these systems are provided in Appendix A.

\subsection{Study Design}

\subsubsection{Participants.} For this exploratory study, we recruited 8 participants who regularly use email for communication, are attentive to tone, and have experience with generative AI tools like ChatGPT and Gemini. The study lasted ~1 hour, and each participant received a supermarket voucher valued 6.4 USD as compensation.

\subsubsection{Writing Tasks.} As an exploratory study, our goal was to surface the range of cognitive and emotional challenges that arise when people use LLMs to craft tone-sensitive messages. To ground the tasks in realistic situations, we conducted a literature review on difficult conversations and compiled representative scenarios across workplace \cite{varga2021makes}, educational \cite{graham2024difficult}, and family/personal settings \cite{keating2013family}. These scenarios captured common yet emotionally charged situations such as giving critical feedback, mediating peer conflict, requesting support from supervisors, and setting boundaries with relatives.

We refined this collection by removing topics that were either too sensitive for research purposes or unlikely to resonate with our participant pool (e.g., healthcare or elder-care situations). The result was a curated set of 16 tasks (see Appendix E) that varied in relational closeness, emotional stakes, and communicative goals. Participants selected the tasks that felt personally relevant and could also contribute their own challenging situations. This open-ended approach emphasized authentic reflection and spontaneous tone discovery rather than standardized performance.

\subsubsection{Procedure.} Each session lasted approximately one hour. Participants used all four systems with think-aloud protocols and brief semi-structured interviews before and after each task. This process helped us capture their intent-articulation strategies, system interactions, and perceived challenges. We analyzed both qualitative observations and light quantitative indicators (e.g., writing time and revision counts) to triangulate patterns described in Section 3.2.

\subsection{Results: Cognitive Challenges in LLM-Assisted Email Composition}

Participants composed 24 emails across 14 of the 16 curated scenarios, covering tasks such as requesting salary increases, negotiating authorship order, and ending collaborations. 

On average, emails were $M = 201$ words in length ($SD = 58$, $range [118, 352]$). Across systems, completion times averaged 8 minutes for Friday Email, 9.9 minutes for Textoshop paired with Gemini, and 10.1 minutes for Gemini with Gmail.

Our analysis revealed a recurring workflow unfolding in three phases: (1) translating goals into effective prompts; (2) evaluating and refining imperfect drafts; and (3) restarting for similar tasks. Each phase surfaced distinct cognitive challenges.

\subsubsection{Challenge 1: Articulating Nuanced Tone Requirements}

A central challenge lies in how users articulate their tonal intentions to the AI. Participants find it challenging to articulate the contextual nuances that shaped their communication goals using prompts. We identified two recurring articulation patterns that shaped how users approached this task.


\textbf{Context-Rich Articulators} (e.g., P1, P2, P5) invested significant effort upfront, composing long, descriptive prompts embedding narrative background and relationship dynamics. They viewed this preparation as essential to “get a good first draft,” since “multi-round dialogue has high costs” (P5) and “iterative prompting does not guarantee linear improvement” (P1). However, articulating complex intent remained demanding. Several described “spending 20 minutes thinking about these nuances” before writing (P3) and feeling “scattered and disorganized, thinking of things as they go” (P3, P8). Despite their efforts, important contextual cues were often omitted— e.g., calling someone a “best friend” without explaining the closeness or emotional history that defined the relationship (P1) leading to misaligned drafts that failed to capture the intended tone.

\textbf{Vague Intent Describers} (e.g., P6, P7) provided brief, general prompts that relied on single tone labels such as “polite” or “casual.” They often realized that these terms failed to communicate their intended nuance, creating a mismatch between their meaning and the AI’s interpretation. This issue was especially evident in predefined systems like Friday Email, where participants found tone categories “too rigid” or “not expressive enough” to reflect their needs (7 of 8 participants).

Across both groups, participants described tone articulation as cognitively taxing, requiring externalization of rarely verbalized thoughts. Several participants expressed a desire for systems that could help them clarify and structure their tone goals before writing. P2, for instance, hoped that “future systems could help users consider all communicative nuances from the outset,” while P8 explained that “if the system could help me clarify my own needs at the beginning, then the work I need to do later is actually just modifying some expressions. At least I would be clear about what I'm thinking.” This recurring theme points to a design opportunity to support reflection and pre-writing planning.

\subsubsection{Challenge 2: Modifications at Different Levels of the Email}
After generating an initial draft, participants faced their second major difficulty—modifying AI output to achieve the desired tone and relational balance. Although current LLMs can produce fluent text, participants frequently described frustration when trying to refine specific parts of a message without losing coherence or tone consistency. 

Most existing systems offer control at two extremes: regenerating entire drafts or editing individual words. Our observations show that writers need something in between—support for mid-level, structured editing that allows selective tone adjustments across sections. P3 described varying emotional emphasis: “My frustration, sadness, and apology—these emotions should be expressed in different parts of the email. I need to target each part to confirm where exactly I express my frustration, where I express my sadness, and where I express my apology.” P7 described wanting “to divide the email into blocks and process different blocks together,” explaining that when persuading a supervisor to fund equipment, “the introduction should feel informal and close, but the justification should sound more formal, and the motivation can stay simple.” These accounts converge on a shared limitation: current systems provide no way to manage tone at a rhetorical-move \cite{faqe2025linguistic} (communicative units) level, leaving users to re-edit entire drafts or stitch together partial revisions manually.



Participants’ calls for modular or section-aware editing highlight a broader design opportunity for LLM systems—to support writers in coordinating tone across distinct communicative functions, not just sentences.

\subsubsection{Challenge 3: Reusing Effective Tone Strategies Across Emails}

An interesting finding emerged when participants spent considerable time fine-tuning their messages—many expressed a desire to reuse their successful modifications in future communications. For instance, P1 noted that "LLMs always open emails with 'I hope this message finds you well,' but I always replace it with my personal greeting style. This preference isn't saved anywhere, and I wish it could be automatically reapplied." Similarly, P4 described systematic patterns in their editing: "I consistently remove verbose explanations and condense them to be more direct. If these editing patterns could be saved, it would save me so much time."

This observation revealed a key insight: many personal preferences, stylistic choices, and situational strategies are actually reusable across similar contexts. However, current systems treat each email as an isolated interaction, requiring users to reconstruct their communicative preferences from scratch each time. This acts as an "internal interruption," \cite{mcfarlane2002scope} forcing users to break their creative flow to recall or rediscover effective strategies for recurring scenarios, which erodes efficiency and dilutes their communicative identity. While commercial tools like MailMaestro support template-based reuse through variable substitution for structural elements (names, dates, addresses), these mechanisms miss the nuanced tonal patterns that define effective interpersonal communication.

Our analysis revealed two distinct levels of reuse that participants desired. First, granular modification patterns—systematic personal preferences in editing behaviors that transfer across tasks, such as replacing generic openings or condensing verbose AI-generated explanations. Second, strategic communication patterns organized around recipient relationships and contexts. As P4 explained: "I write to certain groups of people in similar tones... to colleague B in a similar tone as colleague A, or when seeking help from unfamiliar people, I use consistent approaches."

Without systematic reuse mechanisms that capture both relational dynamics and communicative contexts, users face unnecessary cognitive overhead in their AI-assisted writing. 

The deeper insight from our formative study is that our writing is shaped by many factors we may not always know how to explicitly articulate, yet these factors influence every message we send. Given our limited short-term memory and the cognitive cost of deliberate reflection, it becomes difficult for users to effectively retrieve and communicate all these nuances to AI during prompting. If there were a way to save, store, and organize these factors and preferences into easily manageable operational instructions that can be browsed, selected, and applied, this would allow AI systems to increasingly understand personal needs. This approach transforms hidden and hard-to-describe elements into something much more accessible, creating a powerful complement to current general-purpose AI for communication.


\subsection{Design Goals}

Building on the cognitive challenges identified in our formative study, we propose three design goals that aim to make AI writing systems more reflective, controllable, and personally meaningful to their users.

\subsubsection{Goal 1: Support the Articulation of Communicative Intent}
\label{Design Goal 1}
Users begin email composition with partial intent, knowing their desired relational stance (respect, warmth, firmness) but struggling to translate these feelings into explicit instructions. Rather than demanding complete specifications upfront, systems scaffold tone discovery through guided reflection and iterative clarification, becoming partners in metacognition rather than mere text generators.

\subsubsection{Goal 2: Enable Granular and Hierarchical Control}
\label{Design Goal 2}
Current tools' binary modes (whole-text regeneration or word-level edits) collapse the multi-layered structure of human communication. Users need mid-level control aligned with how they cognitively structure messages—opening, justification, closure. Communicative-Unit-based editing restores compositional authorship while maintaining coherence across emotional and functional sections.

\subsubsection{Goal 3: Instrumentalize Self-Understanding for Future Reuse}
\label{Design Goal 3}
Tone preferences represent stable yet tacit aspects of communicative identity. Systems treat users' past adaptations as self-learning data, capturing individual preferences and transforming them into reusable, context-aware patterns. This shifts LLM assistance from static prompting toward co-adaptive relationships where systems internalize user style and values, amplifying self-understanding rather than replacing authorship.

\vspace{3mm}
These goals reframe AI writing assistance from prompt-and-generate paradigms into exploratory, adaptive processes that foster reflection, expressive control, and identity continuity across contexts (see Table~\ref{tab:persona_mail}).

\section{PersonaMail}
Based on our formative study findings, PersonaMail \footnote{The source code for PersonaMail is available at \url{https://github.com/Synteraction-Lab/IUI26-PersonaMail}.}  was designed to address the discovered challenges. In this section, we detail how PersonaMail's design translates these insights into functional features that directly address user needs identified in our formative research.


\begin{figure*}[t]
  \centering
  \includegraphics[width=\textwidth]{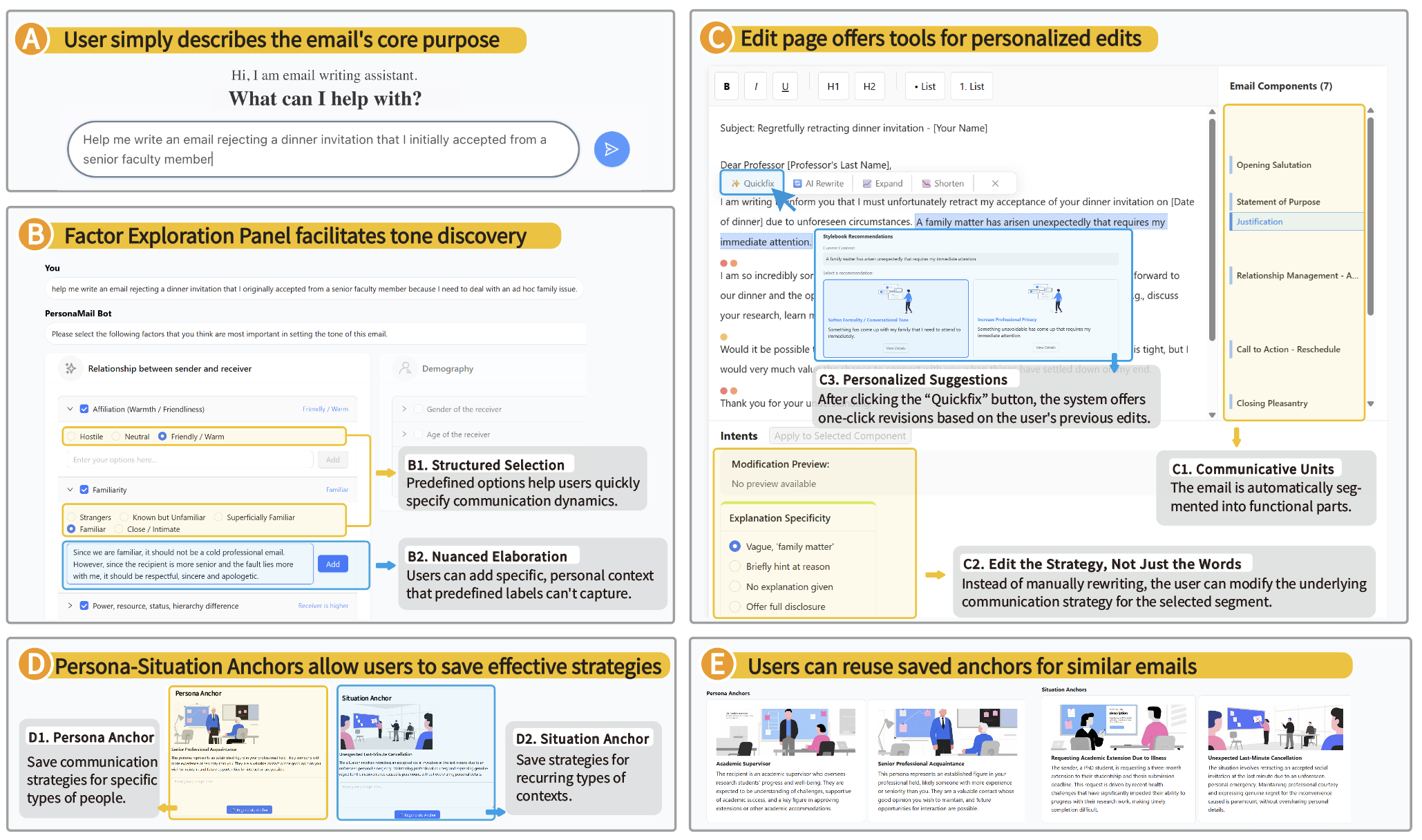}
  \caption{System overview of PersonaMail showing the integrated architecture for tone articulation and email composition. \textbf{(A)} In the task prompt stage, users simply describe the email's core purpose without specifying tone. \textbf{(B)} The Factor Exploration Panel scaffolds users' reflection on research-based communicative factors through \textbf{(B1)} structured quick selection and \textbf{(B2)} open-ended elaboration. \textbf{(C)} The text-edit interface enables granular tone adjustments through communicative-unit-based editing. The email body is segmented into labeled components such as Opening Salutation, Justification, and Closing Pleasantry \textbf{(C1)}. When users select a segment, its associated communicative intents \textbf{(C2)}—such as Explanation Specificity, Formality Level, and Emphasis on Regret. Users can preview and apply alternative strategic variations (e.g., shifting from vague reason to full disclosure), observing immediate text-level updates. The Adaptive Stylebook's Quick Fix interface \textbf{(C3)} surfaces personalized revision patterns learned from user edits. When users select text, the system presents alternative phrasings that reflect previously observed stylistic tendencies (e.g., softening tone or increasing privacy), transforming past edits into adaptive, reusable communication instruments. \textbf{(D)} Persona-Situation Anchors allow users to save successful communication strategies. The system enables creating anchors based on recipient relationships (Persona, D1) or communication contexts (Situation, D2), with AI-generated descriptive names that users can refine to match their reuse intentions. \textbf{(E)} In future writing stages, users can apply saved anchors to pre-configure the system's communicative factors, generating an initial draft that mirrors their preferred tone expectations with minimum effort.}
  \Description{Overview diagram of PersonaMail's system architecture illustrating the relationship between factor exploration, rhetorical move segmentation, and intent-driven modification components.}
  \label{fig:systemfigure}
\end{figure*}

\subsection{Supporting Tone Articulation through Curated Factor Scaffolds (\textit{Design Goal 1})}


To systematically identify factors that shape tone in interpersonal communication, we conducted a literature review across communication and social psychology research.

\subsubsection{Scope}   

Although the term \textit{tone} is widely used to describe affective variations in written communication, definitions across disciplines differ, ranging from linguistic politeness markers in pragmatics to emotional stance and relational alignment in social psychology \cite{mccawley1971concept}. 
To ensure comprehensive coverage, our review intentionally broadened its scope beyond "tone" per se to include diverse communication delivery features reflect the practical surface layer through which tone is operationalized in interpersonal and professional correspondence.

\subsubsection{Method}
Following PRISMA guidelines ~\cite{page2021prisma}, we conducted a systematic review of 97 papers to identify factors influencing written delivery (detailed methodology in Appendix C). Through inductive coding, we identified two overarching categories: Persona Factors and Situation Factors (Table 1).

\begin{table*}[t]
\centering
\caption{Factors Influencing Tone Selection in Written Communication }
\label{tab:revised_factors_tone}
\begin{tabular}{p{1.5cm} p{3cm} p{7cm} p{3cm}}
\toprule
\textbf{Category} & \textbf{Factor} & \textbf{Description} & \textbf{Source} \\
\midrule
\multirow{7}{*}{\textbf{Persona}} 
& Relationship Type 
  & Writer-recipient relationship (e.g., supervisor/student, colleague, friend) 
  & \cite{baggia_emoticons_2022,dube_beyond_2024, decock2017customer} \\
& Familiarity 
  & Level of familiarity 
  & \cite{lewis_instant_2005,xue_how_2025,mehrabi_boshrabadi_cyber-communic_2016, waldvogel2007greetings} \\
& Power/Status 
  & Hierarchical relationship and status differences 
  & \cite{churchill_effectiveness_2021,schlight_teaching_2020,muir_linguistic_2017, liu2011power} \\
& Gender Dynamics 
  & Gender influence on personalization and formality 
  & \cite{maiz-arevalo_influence_2024,jones_gender_2018, pollard2016preferred, jones2016gender} \\
& Personality Traits 
  & Personality Traits (Introverted, Extroverted, Sensitive) 
  & \cite{brunet_are_2008} \\
& Relationship Needs 
  & Expectations for maintaining/improving relationships 
  & \cite{brody_equity_2015, brody2015equity} \\
& Age 
  & Age consideration of the sender and receiver 
  & \cite{tian_relevance_2025,minich_adolescent_2025, perez2015first, sakai2018social} \\
  & Cultural Context 
  & Cultural differences in directness and formality 
  & \cite{atsawintarangku_investigating_2016,richard_saving_2016,bowman_emojis_2023, van2021person} \\
\midrule
\multirow{7}{*}{\textbf{Situation}} 
& Emotional Intent 
  & Emotions to evoke or avoid in recipient 
  & \cite{baggia_emoticons_2022,smith_adolescents_2014} \\

& Competing Goals 
  & Trade-offs between clarity, politeness, and objectives 
  & \cite{schlight_teaching_2020} \\
& Promptness 
  & Urgent/Non-urgent communication needs 
  & \cite{seidl_strategy_2010,bodie_explaining_2011} \\
& Communication Purpose 
  & The purpose of a communication—such as request, complaint, apology, or rejection 
  & \cite{rygg_openings_2021,pinto_shifting_2019} \\
& Occasion 
  & Formal/Informal occasions 
  & \cite{lasan_expression_2024} \\
& Avoid Negative Consequence 
  & To avoid negative outcomes caused by the communication 
  & \cite{smith_adolescents_2014} \\
\bottomrule
\end{tabular}
\end{table*}

\subsubsection{Operationalization: Factor Exploration and Tone Articulation Panel}
PersonaMail operationalizes these insights through a Factor Exploration Panel that scaffolds users' reflection on communicative nuances (Figure~\ref{fig:systemfigure}-B). After the user enters an initial message context, the system presents a curated subset of relevant factors drawn from the taxonomy. Each factor can be refined through two complementary modes: (1) structured selection, where users choose from system-suggested options (e.g., Familiarity → Familiar) (Figure~\ref{fig:systemfigure}-B1), and (2) open-ended elaboration, where users can qualify or nuance their choice (e.g., "We are familiar, but not close enough to discuss personal matters.") (Figure~\ref{fig:systemfigure}-B2).

\subsection{Communicative-Unit-Based Architecture and Intent-driven Modifications (\textit{Design Goal 2})}



\subsubsection{Communicative Units}
PersonaMail segments emails into functional building blocks called \emph{Communicative Units} (Figure~\ref{fig:systemfigure}-C1), defined by their communicative function. These standardized text modules serve specific purposes in the message structure, such as \emph{Opening\_Salutation}, \emph{Statement\_of\_Purpose}, \emph{Justification}, \emph{Call\_to\_Action}, and \emph{Closing\_Pleasantry}.

\subsubsection{Intents}
To provide both transparent control over the AI's reasoning and granular control over the email's communicative units, PersonaMail implements the principle of reified user intent (Figure~\ref{fig:systemfigure}-C2) \cite{riche_ai-instruments_2025}. This principle turns a user's goal into a tangible, manipulable object. In our system, an intent is an observable and editable "writing instruction"— structured as a [type, value] fragment, such as \emph{[Request Phrasing, Implicit mention]}—that the AI derives from the user's task description and factor selections. 

Intents and communicative units operate through a flexible many-to-many relationship: a single intent can influence multiple units, while an individual unit can be shaped by multiple intents simultaneously. 

Consider an email declining an important dinner invitation from a senior faculty member that you had previously accepted. A single intent, such as \emph{[Relationship\_Preservation, High\_Priority]}, can influence multiple units throughout the message—shaping the \emph{Opening\_Salutation} to convey appropriate respect, the \emph{Justification} to provide context without over-explaining, and the \emph{Closing\_Pleasantry} to reinforce professional rapport. Conversely, a single unit like \emph{Justification} may be shaped by multiple intents simultaneously: \emph{[Excuse\_Strategy, Unavoidable\_Circumstance]} ensures the reason appears legitimate and beyond the writer's control, and \emph{[Relationship\_Preservation, High\_Priority]} maintains the collaborative dynamic.

\subsubsection{Intent-driven Modifications}
To enable granular control that bridges whole-email and word-level editing (Goal 2), we introduce an intent-driven modification mechanism. The generated email appears as standard text, with communicative units shown in a sidebar and associated intents displayed as colored labels below. (Figure~\ref{fig:systemfigure}-C)

By selecting a unit, users view its associated intents and explore alternative expressions by adjusting intent values. This structured navigation prevents users from getting lost in random AI-generated alternatives.

This mechanism is implemented through a pipeline (Figure~\ref{fig:pipeline}, columns 3 and 4) of four specialized prompt-based agents working post-generation: the \textit{Intent Analyzer} derives the potential intent space from user inputs; the \textit{Communicative-Unit Extractor} segments the email functionally; the \textit{Unit-Intent Linker} maps units to intents; and the \textit{Intent-Driven Rewriter} generates alternative text when users modify intent values.

Consider declining a dinner invitation from a senior faculty member. The system externalizes its strategy into editable intents:

\begin{itemize}
  \item \textbf{[Opening Strategy, Apology-first, acknowledgment-focused]}: The system's default choice, with alternatives like ``Direct cancellation notice'' or ``Context-first explanation.''
  \item \textbf{[Excuse Strategy, Unavoidable circumstance, legitimizing]}: The default way to frame the cancellation reason, with alternatives like ``Personal emergency, brief'' or ``Detailed explanation, transparency-focused.''
  \item \textbf{[Relationship Preservation, High priority, future-oriented]}: The default approach to maintaining rapport, with alternatives such as ``Minimal acknowledgment'' or ``Extensive reassurance, status-conscious.''
\end{itemize}

This interface enables users to modify strategic approaches rather than just text, providing clear insight into the AI's reasoning.

\subsection{Communication Strategy Reuse: Persona-Situation Anchors and Adaptive Stylebook (\textit{Design Goal 3})}

To address the challenge of reusing effective communication strategies, we designed a two-tiered reuse system. Persona-Situation Anchors capture high-level communication strategies for similar recipient relationships and writing contexts, while the Adaptive Stylebook learns user modification preferences at the text level.

\subsubsection{Persona-Situation Anchors}
Persona-Situation Anchors (Figure~\ref{fig:systemfigure}-D) enable users to save and reapply effective communication strategies by capturing the underlying configuration of tone factors that shaped a successful email. Each anchor encapsulates the factor configuration (e.g., familiarity, power distance, emotional intent) that defined a preferred communicative approach.

After completing a challenging message, users can save this configuration as either a Persona or Situation Anchor. For instance, after crafting an apology to a senior academic mentor, the system might suggest saving it as "Familiar Senior Academic Mentors," recording factor selections such as high familiarity, significant power difference, and relationship preservation priority. When later composing an email to another senior collaborator, applying this anchor automatically pre-loads these settings with contextually appropriate adaptations.

The system adapts saved factors to the new context through an LLM-based \textit{Anchor-Adaptation Agent} that operates on semantic transformation principles. When applying an anchor, the agent analyzes the shift between the source configuration (saved factors and original task) and target context (current writing task), determining which factors should remain constant (e.g., Power Distance in similar hierarchies) and which must transform (e.g., Communication Purpose from "Apology" to "Request"). This achieves context-aware reuse while reducing user effort and promoting tonal coherence across a user's personal communication history.

\subsubsection{Adaptive Stylebook}
The Adaptive Stylebook operates at a granular level by building a personalized library of a user's common modification patterns (Figure~\ref{fig:systemfigure}-C3). This mechanism captures user edits along with explanations of why changes were made, creating concrete examples of communication style in action that become reusable templates.

When users edit AI-generated content, they reveal stylistic preferences through their modifications. For instance, when an employee softens a direct budget request to their manager—adding diplomatic language and acknowledging financial constraints—this reveals a preference for relationship-preserving communication in hierarchical workplace contexts. The system saves this pattern as a reusable template for future sensitive requests to management.

This learning process is operationalized through an \textit{Edit-Analysis Agent} that converts implicit behaviors into explicit style rules. When users modify text, the system captures the original content, revised version, and user-inputted rationale (prompted by "Why did you make this change?"). The agent then synthesizes these inputs into a structured Stylebook, including modification name, original text, revised text, modification rationale, receiver description and occasion description. For future tasks, when users activate "QuickFix," a Retrieval Agent scans the selected content against saved records, retrieving relevant patterns based on semantic and contextual similarity to the user's historical preferences, making the implicit explicit without requiring users to describe their writing style abstractly. The result is a personalized writing assistant that becomes more attuned to individual communication patterns over time.

\begin{table*}[t]
\caption{PersonaMail Design Challenges and Solution Approaches}
\label{tab:persona_mail}
\begin{tabular}{p{2.8cm}p{4cm}p{3.5cm}p{4.5cm}}
\toprule
\textbf{Challenge Category} & \textbf{Specific Problem} & \textbf{Design Goal} & \textbf{PersonaMail Solution} \\
\midrule
\multirow{2}{2.8cm}{Articulating Nuanced Tone Requirements} 
& Vague initial intent - users begin with unclear communicative goals 
& \multirow{2}{3.5cm}{Goal 1: Facilitate Communicative Nuances Discovery} 
& Structured exploration scaffolding to help users discover tone preferences through iterative interaction \\
\cmidrule{2-2}\cmidrule{4-4}
& Oversimplified tone categories - predefined labels fail to capture complex needs 
&  
& Support nuanced tone specification beyond categorical limitations through exploratory interface \\
\midrule
Multi-Level Control Needs 
& Limited granularity - tools only support whole-email or word-level editing 
& Goal 2: Enable Granular Control Without Complexity Explosion 
& Enable intervention at multiple hierarchical levels while maintaining usability \\
\midrule
\multirow{2}{2.8cm}{Reusing Effective Tone Strategies} 
& Context-similar scenarios require similar tone strategies 
& \multirow{2}{3.5cm}{Goal 3: Instrumentalize Self-Understanding for Future Efficiency} 
& Organize personalized templates by recipient type (Persona) and communication context (Situation) \\
\cmidrule{2-2}\cmidrule{4-4}
& No way to capture successful modifications 
&  
& Transform user modification patterns into reusable instruments (Adaptive Stylebook) \\
\bottomrule
\end{tabular}
\end{table*}

\subsection{Summary}
\begin{figure}[h]
  \centering
  \includegraphics[width=1\linewidth]{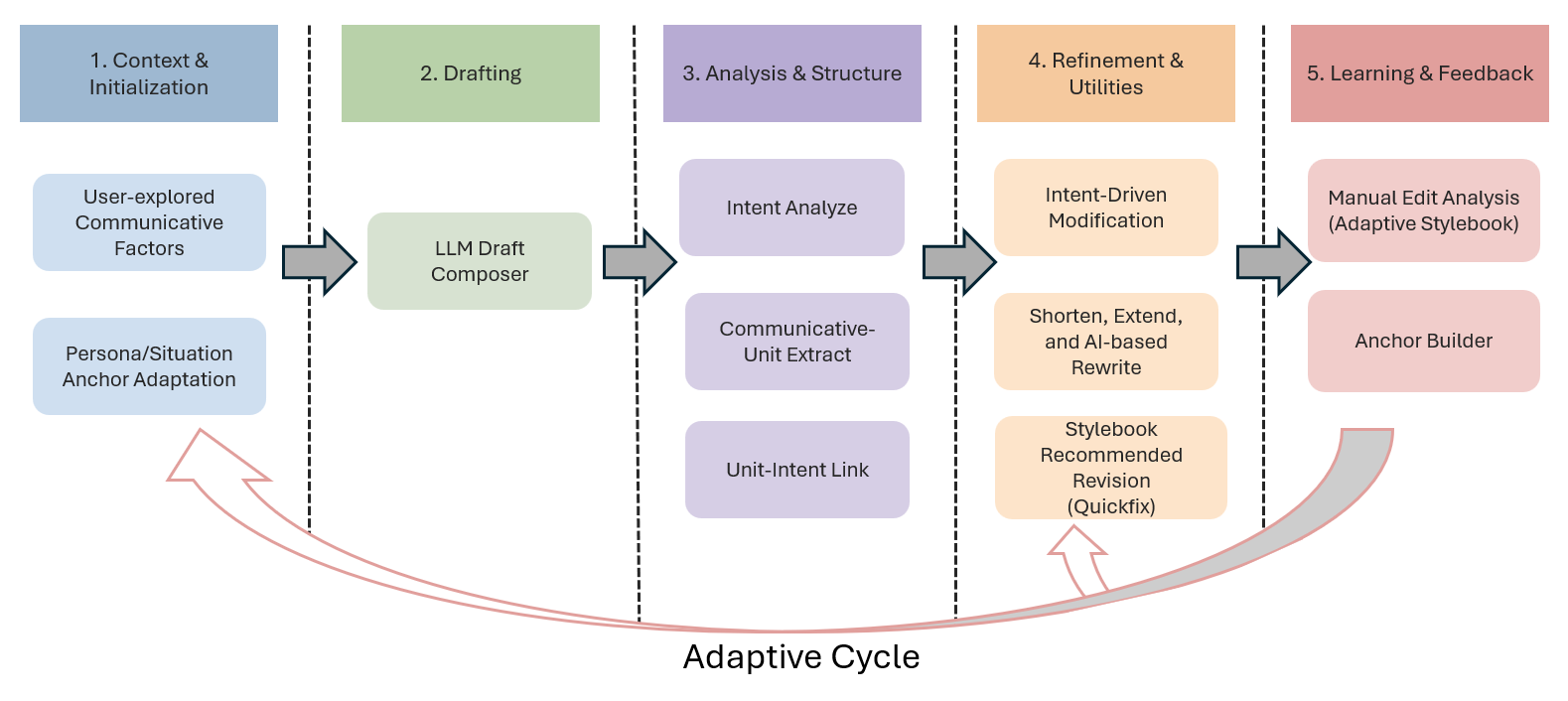}
  \caption{The PersonaMail System Pipeline. The workflow illustrates the information flow from initialization to adaptive learning. (1) \textbf{Context \& Initialization}: User intent is captured via the Factor Exploration Panel and retrieval and adaptation of existing Persona/Situation Anchors. (2) \textbf{Drafting}: The LLM generates a base draft. (3) \textbf{Analysis \& Structure}: The system decomposes text into Communicative Units and links them to identified Intents. (4) \textbf{Refinement \& Utilities}: Users refine the draft using Intent-Driven Modifications, QuickFix, and other AI-based tools. (5) \textbf{Learning \& Feedback}: The system captures manual edits to update the Adaptive Stylebook and allows users to save new Anchors, which feed back as the \textit{Adaptive Cycle} for future tasks.}
  \Description{A five-stage diagram showing the PersonaMail workflow. Stage 1 is Context and Initialization, feeding into Stage 2 Drafting. Stage 3 is Analysis and Structure, breaking the draft into intents and communicative units. Stage 4 is Refinement and Utilities, allowing modifications. Stage 5 is Learning and Feedback, which captures manual edits and anchors. A large arrow labeled Adaptive Cycle connects Stage 5 back to Stage 1 and Stage 4.}
  \label{fig:pipeline}
\end{figure}

PersonaMail integrates the three design goals into a cohesive system that evolves with user interaction, as illustrated in the system pipeline (Figure~\ref{fig:pipeline}). Factor Exploration supports the articulation of nuanced communicative intent (Goal 1), the Communicative-Unit and Intent architecture enables precise, hierarchical tone control (Goal 2), and the combined Persona-Situation Anchors and Adaptive Stylebook facilitate long-term personalization and reuse (Goal 3).

Table \ref{tab:persona_mail} summarizes how each feature directly addresses the cognitive challenges identified in our formative study.
\section{Evaluation}

Building on the design goals established in Section 3.3, we conducted a user study to examine how PersonaMail supports users in composing nuanced, tone-sensitive emails. The study evaluates the system’s three core innovations — factor-based scaffolding, multi-level control, and adaptive personalization - each addressing a cognitive or interactional challenge identified in our formative analysis (Section 3).

This evaluation aims to understand not only whether PersonaMail improves writing outcomes, but how its mechanisms transform the process of human–AI collaboration in communication. Specifically, we investigate how structured tone articulation (Goal 1) influences clarity of intent, how granular control (Goal 2) affects revision efficiency, and how adaptive learning (Goal 3) supports reuse and consistency across tasks. Thus, the following four research questions guide our investigation:

\begin{enumerate}[label= \textbf{RQ\arabic*:}, nosep]
    \item \textbf{Email Quality and Articulation Effectiveness.} Does PersonaMail's factor-based scaffolding enable users to more comprehensively articulate their communicative intent compared to open-ended prompting, and does this enhanced articulation lead to higher-quality email outputs? 
    
    \item \textbf{Control and Revision Efficiency.} How effectively does PersonaMail's multi-level control architecture—combining intent-driven modifications and communicative-unit based editing—support efficient and precise revision compared to traditional text editing workflows? 
    
    \item \textbf{Adaptive Learning and Reusability.} Do PersonaMail's adaptive features—Persona-Situation Anchors and the Adaptive Stylebook—successfully reduce articulation burden and improve efficiency across repeated usage without compromising email quality? 
    
    \item \textbf{Cognitive Load and User Experience.} Does PersonaMail's structured approach to communication assistance reduce the cognitive burden of composing tone-sensitive emails compared to open-ended conversational AI interfaces? 
\end{enumerate}

\subsection{Participants}
We recruited 16 participants (8 female, 8 male), aged 23-32 (M = 26, SD = 3), from diverse backgrounds such as design, computer science, biomedical science, chemistry, and law. Most were students or early-career professionals with 0-6 (M = 2.28, SD = 1.84) years of working experience who had used email as their primary communication medium (M = 5 years, SD = 3). All participants self-reported familiarity with conversational large language model interfaces such as ChatGPT or Gemini. Each session lasted approximately 2 hours, and participants received \$12.85 USD as compensation for their time. All procedures were approved by the university's Institutional Review Board (IRB).

\subsection{Comparison Systems}
To evaluate PersonaMail’s effectiveness, we compared it with a representative baseline system reflecting current AI-assisted email writing practices. Both systems used the same underlying LLM (Gemini 2.5 Flash) to control for model effects.

\textbf{Baseline.} The baseline combined a Gemini chat interface with Gmail editing, representing the prevalent two-stage workflow in which users first craft prompts conversationally with an AI and then refine outputs manually within a standard email editor. This setup captures the dominant pattern of AI generation followed by human revision in everyday writing.

We selected this baseline instead of specialized tools from our formative study (MailMaestro, Friday Email, Textoshop) for two reasons. First, PersonaMail introduces fundamentally new capabilities that extend beyond the existing tools' design space (Table~\ref{tab:designspace} in Appendix B), making comparison with tools with known limitations less informative. Second, the Gemini–Gmail workflow represents a functional superset of specialized tools like MailMaestro that rely on predefined tone labels—an approach our formative study found problematic for challenging conversations. To our knowledge, it offers the closest overall functionality compared with PersonaMail. Thus, we believe the Gemini-Gmail workflow is the more appropriate baseline compared with those previously mentioned specialized tools.

\subsection{Procedure}
Each study session lasted \(\sim2\)hours and followed four sequential phases: briefing, task selection, main task execution, and debriefing. This structure enabled both controlled comparison and observation of adaptive learning effects across repeated use.

\begin{enumerate}[leftmargin=*]
    \item \textbf{Briefing.} Participants received an overview of the study, provided written consent, and completed a demographic questionnaire.

    \item \textbf{Task Selection (Real-World Protocol).} Consistent with the approach used in our formative study (Section~\ref{formativestudy}), participants selected tasks from our curated set of challenging communication scenarios spanning workplace, educational, and personal contexts. They were also encouraged to bring their own difficult writing situations. This design ensured participants engaged with scenarios they could genuinely relate to, fostering authentic tone articulation and emotional reasoning rather than artificial compliance with pre-assigned prompts.

    \item \textbf{Main Task Execution.} The core study comprised three rounds of writing tasks designed to evaluate both comparative performance and PersonaMail’s adaptive learning capabilities:
    \begin{itemize}[leftmargin=1em]
        \item \textbf{Round 1: Baseline vs. PersonaMail Comparison.} Participants used two systems in fixed order: first the baseline (Gemini~+~Gmail), then PersonaMail. The baseline system represents a familiar workflow and thus does not bias subsequent PersonaMail use, whereas exposure to PersonaMail’s factor scaffolds could unfairly improve baseline prompting.
        \item \textbf{Rounds 2 and 3: Adaptive Use of PersonaMail.} Participants then composed two additional emails using PersonaMail exclusively, allowing evaluation of the system’s adaptive features (\textit{Adaptive Stylebook} and \textit{Persona-Situation Anchors}). The three PersonaMail tasks were thematically related, sharing similar relationship dynamics or situational challenges, to enable meaningful assessment of reusability. For instance, one participant wrote three refusals involving close contacts (declining free design work, setting financial boundaries with a sibling, and refusing a referral request), each requiring similar tone management strategies. Task sets were validated by experts and iteratively refined by the authors until reaching consensus; representative tasks are provided in Appendix~F-1, and full validation procedures are detailed in Appendix~F-2.
         
    \end{itemize}

    \item \textbf{Debriefing and Interview.} After completing the writing tasks, we conducted a semi-structured interview to reflect on participant experience, cognitive strategies, and perceptions of PersonaMail’s support mechanisms.
\end{enumerate}

\subsection{Data Collection}

We adopted a mixed-methods approach combining quantitative performance metrics and qualitative interview data to evaluate PersonaMail.

\subsubsection{Quantitative Data}

We collected comprehensive quantitative metrics addressing our four research questions. Table~\ref{tab:metrics} summarizes all measures, which were manually coded from recorded sessions or collected through questionnaires. Detailed descriptions of these quantitative metrics are provided in Appendix~D.

\begin{table*}[t]
\centering
\caption{Quantitative metrics collected to address research questions}
\label{tab:metrics}
\small
\setlength{\tabcolsep}{4pt}
\renewcommand{\arraystretch}{0.9}
\begin{tabular}{p{0.15\textwidth}p{0.35\textwidth}p{0.42\textwidth}}
\toprule
\textbf{RQ} & \textbf{Metric Category} & \textbf{Specific Measures} \\
\midrule
\textbf{RQ1: Email Quality and Articulation Metrics} & Email Quality & First draft quality, revised draft quality (7-point Likert scale) \\
\cmidrule(lr){2-3}
& Articulation Depth & \textit{PersonaMail}: Number of factors engaged, total word count (prompt~+~factors), articulation time. \textit{Baseline}: Number of aspects mentioned (manually coded), prompt word count, prompting time \\
\midrule
\textbf{RQ2: Control and Revision Efficiency Metrics} & Revision Efficiency & Revision time for both systems \\
\cmidrule(lr){2-3}
& Control Mechanisms & \textit{PersonaMail}: Distribution across intent modifications, Quick Fix, text editing; acceptance rates for suggestions. \textit{Baseline}: Distribution across conversational prompting and manual editing \\
\midrule
\textbf{RQ3: Adaptive Learning and Reusability Metrics} & Adaptive Reusability & \textit{PersonaMail}: Factors engaged per round, factors revised when using anchors, temporal metrics (prompting time, revision time, total time) across Rounds~1--3, quality ratings across rounds \\
\midrule
\textbf{RQ4: Cognitive Load Metrics} & Cognitive Load & NASA-TLX Raw score~\cite{hart1988development, hertzum2021reference} (10-point scale; higher~=~greater load) for both systems \\
\bottomrule
\end{tabular}
\end{table*}

\subsubsection{Qualitative Data}

To complement the quantitative measures, we collected qualitative data through concurrent and retrospective methods.

\textbf{Think-Aloud Protocols}: During all composition tasks, participants provided concurrent verbal reports of their thought processes, decision-making rationales, and system interactions. These protocols were audio-recorded and later transcribed.

\textbf{Semi-Structured Interviews}: We conducted targeted interviews after each round. Round 1 interviews focused on cross-system comparisons, while Rounds 2-3 interviews explored adaptive learning effectiveness, personalization, and perceived efficiency changes. 

Qualitative data were transcribed verbatim and analyzed using thematic analysis~\cite{braun2006using}. We combined inductive coding (emergent patterns in user experience) with a deductive lens guided by the system’s three evaluation dimensions (articulation, control, reusability).

\section{Results}

We present results across three evaluation dimensions — email quality, efficiency, and cognitive load — each integrating quantitative metrics with corresponding qualitative insights from think-aloud sessions and interviews.

\begin{figure*}[t]
    \centering
    \includegraphics[width=\textwidth]{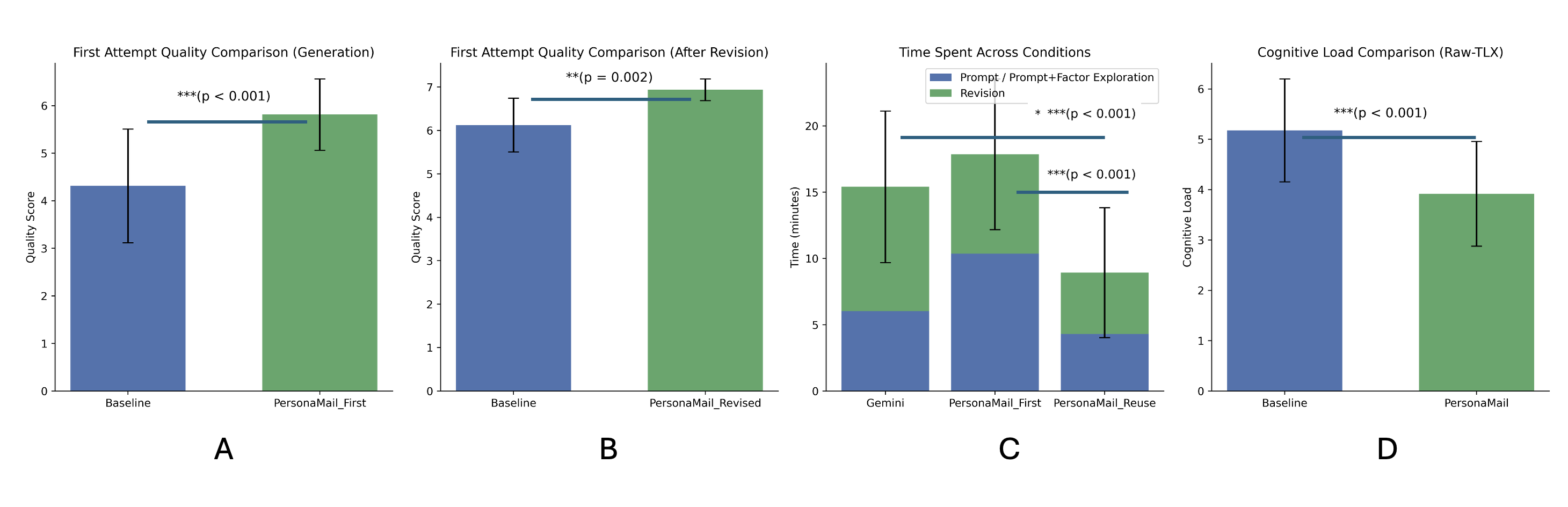}
    \caption{Quantitative evaluation results comparing PersonaMail with the baseline (Gemini+Gmail). PersonaMail significantly outperformed the baseline across email quality, overall efficiency, and cognitive load. (A) First draft quality: PersonaMail's initial drafts were rated 34.8\% higher in quality than the baseline's (p < 0.001). (B) Revised draft quality: After revision, PersonaMail emails maintained significantly higher quality ratings (p = 0.002). (C) Time efficiency: While PersonaMail required more upfront articulation time in the first use, its adaptive features led to a 42.0\% reduction in total task time upon reuse compared to the baseline, and a 50\% reduction compared to its first usage (p < 0.001). (D) Cognitive load: PersonaMail reduced users' cognitive burden by 24.5\% compared to the baseline, as measured by Raw-TLX scores (p < 0.001).  Error bars represent standard deviation. ***p < 0.001, **p < 0.01.}
    \Description{Four bar charts showing comparison between PersonaMail and baseline systems. Chart A shows first draft quality scores with baseline at 4.31 and PersonaMail\_First at 5.81 on a 7-point scale. Chart B displays revised draft quality with baseline at 6.12 and PersonaMail\_Revised at 6.94. Chart C presents a stacked bar chart of time spent across conditions, showing Gemini at approximately 15 minutes, PersonaMail\_First at 18 minutes, and PersonaMail\_Reuse at 9 minutes, divided into prompt/exploration time and revision time. Chart D shows cognitive load with baseline at 5.18 and PersonaMail at 3.91 on Raw-TLX scale. All comparisons show statistical significance markers.}
    \label{fig:datavisualization}
\end{figure*}

\subsection{Email Quality: Enhanced Articulation Leads to Superior Output}

PersonaMail's structured approach to tone articulation significantly improved email quality compared to the baseline system. We assessed quality at two points: immediately after initial AI generation and after user revision.

\subsubsection{First Draft Quality}

The quality of AI-generated first drafts showed substantial improvement with PersonaMail. Users rated PersonaMail's initial output significantly higher (M = 5.81/7, SD = 0.75) compared to Gemini baseline (M = 4.31/7, SD = 1.20), representing a 34.8\% improvement (Figure~\ref{fig:datavisualization}-A). A Wilcoxon signed-rank test confirmed this difference was statistically significant (W = 5.5, p < 0.001, Z = -3.23, r = 0.81), indicating a large effect size.

This quality improvement stems from PersonaMail's structured factor exploration, which enabled more comprehensive and nuanced articulation of communicative factors. While baseline prompts averaged 112.37 words (SD = 50.64), PersonaMail's combined prompt and factor exploration totaled 172 words (SD = 63.65) for the first email—a 53.1\% increase in articulation length. More importantly, PersonaMail users engaged with substantially more communication factors (M = 10.87, SD = 3.09) compared to the free-form baseline prompts (M = 2.62, SD = 1.14), representing a 315\% increase in the breadth of considerations.

Participants attributed quality improvement to PersonaMail's ability to encourage more proactive and comprehensive consideration of communication needs. As P13 observed, "Traditional AI requires users to be good articulators," a limitation that often leads users to "type whatever comes to mind without thorough consideration, merely listing vague and simplistic requirements." (11 out of 16 participants). In contrast, PersonaMail "makes users aware of aspects they would struggle to consider on their own" (P4, P7, P15) and "inspires users to think more comprehensively and proactively" (P2, P5, P9, P13). For instance, P12 described a delicate roommate communication scenario: "when discussing cleanliness in shared spaces with a roommate, you need to strike a balance: acknowledging that you're not very close while maintaining a friendly tone, yet avoiding excessive politeness on crucial points. This nuanced approach to tone can only be achieved using PersonaMail prompts." Overall, such thorough reflection process "enabled the AI to gain deeper contextual understanding, ultimately leading to higher-quality draft generation" (P1, P2, P8, P9).

\subsubsection{Revised Draft Quality}

After user revision, quality ratings remained significantly higher for PersonaMail (Figure~\ref{fig:datavisualization}-B). The revised PersonaMail emails achieved near-ceiling ratings (M = 6.94 out of 7, SD = 0.25), significantly exceeding revised Gemini emails (M = 6.12 out of 7, SD = 0.62). Wilcoxon signed-rank test showed this difference was statistically significant (W = 0.0, p = 0.002, Z = -3.52, r = 0.88).

The superior revised quality reflected not just better initial drafts, but also more effective revision processes enabled by PersonaMail's granular control mechanisms, as discussed in the efficiency section below.

\subsubsection{Quality Consistency Across Repeated Use}

Importantly, email quality remained consistently high across PersonaMail's three usage rounds despite significant efficiency gains. A Friedman test found no significant differences in user preference ratings across the three PersonaMail emails ($\chi^2$(2) = 1.38, p = 0.50), indicating that the system's adaptive features enhanced efficiency without compromising output quality.

\subsection{Efficiency: Adaptive Learning Drives Progressive Time Savings}

PersonaMail demonstrated substantial efficiency gains across multiple dimensions (both task description and draft revision) of the email composition process, with particularly improvements emerging through repeated use as the system's adaptive features accumulated user preferences (Figure~\ref{fig:datavisualization}-C).

\subsubsection{Initial Composition Efficiency}

For first-time use, PersonaMail required greater upfront investment in articulation compared to the baseline. Users spent an average of 10 minutes 21 seconds (SD = 3 minutes 58 seconds) on prompting and factor exploration with PersonaMail versus 6 minutes 2 seconds (SD = 3 minutes 21 seconds) for baseline prompting—a 71.5\% increase. This initial time investment reflects PersonaMail's more comprehensive approach to intent articulation, with users engaging with substantially more communication factors compared to baseline.

However, this upfront investment yielded significant downstream benefits. Revision time for PersonaMail's first draft averaged 7 minutes 30 seconds (SD = 3 minutes 43 seconds) compared to 9 minutes 22 seconds (SD = 4 minutes 18 seconds) for Gemini, representing a 19.9\% time savings. This efficiency gain reflects the higher quality of PersonaMail's initial output, requiring less extensive modification to reach satisfactory quality.

Participants explicitly recognized this efficiency trade-off as worthwhile. Consistent with findings from our formative study (Section 3.2.1), participants favored investing more time in comprehensive initial prompting to obtain a high-quality first draft, rather than extensively revising a suboptimal one. Unlike PersonaMail, the baseline's open-ended prompting often produced drafts where users "need several rounds to achieve a satisfactory result" (P1) or where "The generated emails may overlook certain important aspects or produce messages that leave users at a loss as to how to modify them" (P4, P5, P8, P9). 

\subsubsection{Substantial Efficiency Gains Through Adaptive Reuse}

The true efficiency advantages of PersonaMail emerged in the second and third email compositions, where adaptive features enabled dramatic time savings. Prompting and factor exploration time dropped to just 4 minutes 18 seconds (SD = 2 minutes 6 seconds) for emails utilizing Persona-Situation Anchors—a 58.5\% reduction compared to PersonaMail's first use and a 28.7\% reduction compared to baseline prompting time.

This efficiency gain resulted from anchors pre-populating factors based on previous successful communications. Users still engaged with a rich set of communication factors (M = 11.15, SD = 3.04), but needed to make only minimal adjustments (M = 1.59 factors revised, SD = 2.18) rather than articulating their entire communicative intent from scratch. 

Revision time similarly decreased to 4 minutes 37 seconds (SD = 3 minutes 27 seconds) in reuse conditions—a 38.4\% reduction compared to PersonaMail's first use and 50.7\% faster than baseline revision. This acceleration stemmed from two sources: increased familiarity with PersonaMail's revision techniques after first use, and the Adaptive Stylebook's accumulated Quick Fix records, which enabled one-click application of personalized modification patterns learned from previous emails.

Notably, 33.67\% of all revisions in reuse conditions were accomplished through Quick Fix—simple click-based applications of previously learned preferences, such as "make the tone straightforward and clear," "don't be so sappy," "highlight shared benefits," "don't be too humble," and "remove the pleasantries." etc.

P9 described the progressive efficiency evolution: "From the first to the third time, it became increasingly smooth... I found that most things didn't need modification after applying anchors and transferring factors, and small issues could be fixed with a quick click of Quick Fix—very efficient." 

\subsubsection{Overall Process Efficiency}

When examining total task time (prompting + revision), the efficiency advantages of PersonaMail's adaptive approach became even more pronounced. While first-time PersonaMail use required comparable total time to baseline, reuse conditions achieved remarkable gains. The average total time for reused PersonaMail conditions was 8 min 55 seconds compared to 15 min 24 seconds for Gemini baseline—a 42.0\% reduction in overall task time (Figure~\ref{fig:datavisualization}-C). 

P14's experience captured this progression: "The first time I answered questions very carefully, the second time I only needed to quickly verify, the third time I increasingly trusted that the system could give more accurate output."

\subsubsection{Granular Control Mechanisms Support Efficient Revision}

PersonaMail's multi-level control architecture contributed to revision efficiency. The intent-driven modification mechanism enabled strategic rather than surface-level adjustments. Users averaged 28.27\% (SD = 30.24\%) of their operations through intent modifications, which paired with Quick Fix achieved 61.94\% (SD = 29.76\%) of revisions utilizing PersonaMail's innovative control mechanisms. The remaining 38.06\% comprised traditional text editing and standard AI tools (expand, shorten, localized prompting), indicating that PersonaMail's specialized features handled the majority of revision needs while maintaining compatibility with conventional editing approaches when needed.

The high acceptance rates for these features—93.62\% (SD = 14.88\%) for intent modifications and 88.63\% (SD = 30.59\%) for Quick Fix—indicate that the system's suggestions aligned well with user intentions, reducing the cognitive burden of revision.

Users appreciated Intent's ability to provide different writing styles for comparison, enabling quick identification of preferred versions (P2, P3, P12). This mechanism improved efficiency by eliminating the dual burden of self-diagnosis and articulation. Intent proved especially valuable for users experiencing vague dissatisfaction with generated text but lacking clear revision strategies. As P4 explained, "If I need to modify it by myself, I usually need to first think about what my intent really is," followed by the challenge of "how to use concrete language to express my meaning." Rather than becoming trapped in unproductive deliberation, users received concrete alternatives that allowed them to quickly navigate to preferred versions. P5, P10, P13, and P16 similarly noted: "It's extremely helpful for passages that feel awkward to read but you're unsure how to revise. It essentially offers a more multidimensional interpretation of a sentence, providing users with more suggestions." The efficiency gains were particularly pronounced when system suggestions aligned with user needs: "when the intent suggested options highly align with my modification needs, it is very fast to iterate between different writing strategies" (P7, P11).

\subsection{Cognitive Load: Structured Scaffolding Reduces Mental Burden}

While PersonaMail's comprehensive factor exploration required greater upfront time investment, it paradoxically reduced rather than increased cognitive load. A paired t-test revealed that participants experienced substantially lower cognitive load with PersonaMail (M = 3.91, SD = 1.03) compared to the Gemini baseline (M = 5.18, SD = 1.02), t(15) = 4.87, p < 0.001, Cohen's d = 1.22, representing a very large effect size and a 24.5\% reduction in reported cognitive load (Figure~\ref{fig:datavisualization}-D). This counterintuitive finding suggests that structured scaffolding, despite appearing more complex, actually reduces mental burden by transforming open-ended generation tasks into guided selection processes.

\subsubsection{Reducing Articulation Burden Through Structured Exploration}

The baseline system's open-ended prompting created substantial cognitive burden. P10 explained: "The cognitive load of traditional interfaces is not low, because the generated text is often not very good and requires prompt adjustment." Similarly, P11 and P12 noted that effective baseline use required "describing many details, very detailed prompts, no different from writing it yourself, the process is very hard, the person mainly writing the email is still yourself, workload is very high."

In contrast, PersonaMail's factor-based scaffolding transformed articulation from an open-ended generation task to a structured selection task. P3 and P13 appreciated that "No need to think too much throughout the process, especially when prompting you only need to type the task without thinking about how to teach AI to write, and with factors you only need to look at the pre-provided suggestions and choose (in most cases), much simpler."

Similarly, P6, P8, P10, P11, and P15 emphasized that factors "reduced the workload of thinking, the preset options are also very comprehensive, reducing the time for thinking about prompts and typing, very convenient."

\subsubsection{Quick Fix: Capturing Nuanced Preferences Without Articulation}

The Adaptive Stylebook's Quick Fix feature particularly exemplified PersonaMail's approach to reducing cognitive load by learning from demonstrated preferences rather than requiring explicit articulation. P8 described its value for capturing subtle stylistic preferences that resist verbal description: "when I creates a record that makes language more 'concise,' it's very difficult to directly describe this modification intent in words, because 'concise' is an abstract language feeling, not simply 'shorter.'" For instance, when P8 was writing an email requesting a 3-month studentship extension from their supervisor, they adjusted the AI-generated draft to avoid overly appreciative language about the supervisor's mentorship, creating a Quick Fix record for "concisely expressing appreciation while avoiding excessive tone." This nuanced preference—distinguishing between appropriate gratitude and obsequious flattery—would be difficult to articulate as an explicit instruction, yet Quick Fix successfully captured and reapplied this subtle tonal calibration in subsequent emails.

P15 similarly noted that Quick Fix "very well captured modification intentions that users want to express but find difficult to express."

This capability to operationalize tacit preferences without requiring explicit articulation represents a fundamental advantage over traditional systems that demand users translate all preferences into natural language instructions.

\section{Discussion}

In this section, we reflect on the broader implications of our findings for AI-assisted communication systems. We begin by examining how PersonaMail's design philosophy operationalizes communication theory through its integrated features. We then discuss critical privacy and ethical considerations that emerge from personalized tone learning. Finally, we acknowledge the study's limitations and outline directions for future research.

\subsection{Design Philosophy: Operationalizing Communication Theory Through Integrated Features}

PersonaMail's architecture grounds AI assistance in established communication theories. Rather than treating tone adjustment as a purely linguistic challenge, the system operationalizes theoretical frameworks from interpersonal communication research into concrete design features. Its three core features work in tandem to bridge intent, control, and continuity while embodying distinct theoretical perspectives.

    \subsubsection{Factor-Based Scaffolding for Intent} Drawing on communication research, PersonaMail translates tacit social knowledge into explicit, adjustable factors (e.g., relationship closeness, emotional stakes). This scaffolding helps users externalize nuanced goals (e.g., "respectful but firm with a supervisor") without requiring them to articulate every unspoken detail. By making social context tangible, it reduces the cognitive load of prompt engineering and ensures the AI aligns with relational objectives. This operationalizes the Hyperpersonal Model~\cite{walther1996computer}, which theorizes that asynchronous communication channels—by removing the pressure for real-time replies—afford users opportunities for both selective self-presentation (crafting an optimized image of oneself) and deliberate message refinement. PersonaMail's factor panel provides a structured interface for this process, allowing users to manipulate impression management goals as tangible variables, ensuring AI-generated outputs align with intended relational objectives.
    
    \subsubsection{Communicative-Unit and Intent-Driven Control}  PersonaMail enables granular tone adjustments at the level of communicative units by segmenting emails into functional sections (e.g., Opening\_Salutation, Justification) and exposing the AI's underlying reasoning as editable intents. Through this intent-driven control of communicative units, users can manipulate the social function of message segments without direct text editing. This design validates an approach grounded in Speech Act Theory \cite{searle1969speech}. PersonaMail treats each unit as a distinct illocutionary act—the underlying social function of an utterance (e.g., requesting or apologizing). By allowing users to manipulate the illocutionary force (the intensity of the action) independently of the locutionary act (the surface text), the system restores agency over the social action of the message and bridges the gap between human intent and AI generation. 
    
    \subsubsection{Adaptive Reuse for Continuity} Unlike template-based tools, PersonaMail captures successful strategies (e.g., how a user balances warmth and professionalism with peers) as reusable "anchors." These anchors preserve not just content but the underlying social logic, allowing users to apply proven approaches to new but similar scenarios. This reduces redundant effort and maintains consistency in personal style—countering the homogenization risk of generic LLMs. This capability supports Self-Verification Theory \cite{swann2012self}, which posits that individuals strive for coherence in how they are perceived by others to maintain a stable sense of self. By using Anchors to enforce consistent tonal strategies across recurring interactions, users prevent the AI from generating "out-of-character" messages, thereby ensuring their professional identity remains authentic and recognizable over time.

By integrating these features, PersonaMail introduces a novel dynamic, two-phase pacing model to the writing workflow, which extends traditional pacing theory \cite{dix1992pace} in HCI. Phase one employs "slow technology" \cite{hallnas2001slow} principles through the Factor Exploration Panel, using friction to enforce reflection and intent clarification. However, once a strategy proves successful, the system transitions to higher-velocity execution via Anchors. This design also echoes Rhythm Theory \cite{reddy2002finger}, which recognizes that interpersonal communication is not a series of isolated tasks but consists of recurring scenarios; the system essentially treats a user's past communicative strategies as resources for future coordination, reducing the cognitive load of "reinventing" social nuance.

\subsection{Privacy and Ethical Considerations in Personalized Tone Learning}

While PersonaMail's adaptive learning mechanisms offer significant benefits for communication efficiency and personalization, they also raise privacy and ethical considerations that warrant careful attention in design and deployment.

\subsubsection{Data Storage, Privacy, and User Control}

PersonaMail's learning architecture necessarily stores detailed records of users' communication patterns across different contexts, which may introduce novel privacy vulnerabilities beyond traditional email storage. While conventional email clients store only the final messages, PersonaMail captures the underlying social reasoning—encoding users' relational anxieties, power perceptions, and strategic communication tactics as structured, reusable patterns. When users configure messages with settings like "High Risk Aversion" for a "Senior Faculty Member" and save them as anchors, they create what we term a "social graph of vulnerability" \cite{zheleva2011privacy}: a pre-labeled dataset revealing whom they fear, seek to impress, or treat casually, along with explicit strategies for managing these relationships. Unlike standard email archives requiring interpretation, these semantically rich records make users' relational reasoning directly queryable, creating semantic rather than purely statistical disclosure vulnerabilities \cite{batet2018semantic}. The exposure of such social graph data introduces significant security risks, as adversaries could exploit knowledge of users' communication anxieties and relationship dynamics to craft highly targeted social engineering attacks or fraud schemes that manipulate documented vulnerabilities.

To address these concerns, real-world deployment would require robust technical safeguards including end-to-end encryption of stored patterns, multi-factor authentication, and granular user controls for data review and deletion \cite{li2011survey}. Clear transparency mechanisms must communicate what data is stored, how it is used, and who can access it to maintain user trust and informed consent \cite{rossi2020transparency}.

\subsubsection{Risk of Reinforcing Constrained Communication Patterns}

A critical ethical consideration is that adaptive learning systems may inadvertently reinforce communication patterns that limit users' expressive range. This aligns with the ethical framework proposed by Isop to "preserve natural human presence" \cite{isop2025conceptual}. For instance, users who consistently employ overly apologetic language when communicating with senior colleagues—perhaps due to workplace power imbalances or internalized hierarchical norms—may find these patterns captured and reified by the system. Rather than challenging such behaviors, the current design of PersonaMail could normalize and accelerate their reproduction, turning them into cognitive shortcuts that bypass critical reflection. While these strategies may prove effective in achieving short-term communication goals, their automated reuse may prevent users from recognizing opportunities to renegotiate relational boundaries or advocate for more equitable communication norms, potentially undermining long-term communicative agency and professional development.

These risks highlight the need for \textit{reflective personalization}—adaptive systems that prompt periodic reconsideration of learned patterns, aligning with the principles of reflective design which calls for technology to support users in critically examining the values and assumptions embedded within systems \cite{10.1145/1094562.1094569}. Design interventions could include review prompts for user reassessment, contextual warnings when applying patterns across different power dynamics, and transparency features revealing communication asymmetries. Future iterations could shift from passive learning to active dialogue, prompting users to evaluate whether automated patterns reflect deliberate choices or mental rigidity.

\subsection{Limitations and Future Directions}
While PersonaMail demonstrates a promising direction for co-adaptive writing support, we acknowledge several limitations that also highlight important avenues for future research.

First, our evaluation involved a controlled lab study with participants primarily from student and early-career professional backgrounds. While this demographic is well-suited for the communication scenarios we tested, future work should examine how the findings extend to other demographics, such as senior leadership contexts where communication dynamics differ. Additionally, as with any novel prototype study, the Hawthorne effect may influence ratings due to novelty or observation awareness. Future work involving longitudinal field deployments across diverse user populations would help further isolate these effects and validate real-world adoption patterns.

Second, our lab-based study employed thematically clustered tasks to enable controlled evaluation of PersonaMail's reuse mechanisms. While this design was methodologically necessary to isolate adaptive features and to address practical challenges of in-the-wild studies (including the prevalence of routine informational emails that need minimal tone considerations and ethical concerns around sensitive communications), we acknowledge that real-world email contexts are often more heterogeneous. This could reduce the immediate applicability of saved patterns when users frequently switch between vastly different communication scenarios. Longitudinal field studies would provide valuable insights into the system's utility across more diverse email streams.

Finally, while PersonaMail's factor-based scaffolding theoretically provides a foundation for cross-cultural tone adaptation, our study did not empirically validate this capability in multicultural settings. Existing research has documented that LLMs exhibit systematic "WEIRD" biases (Western, Educated, Industrialized, Rich, Democratic) \cite{zhou2025should}, often defaulting to communication norms aligned with Western cultural frameworks. PersonaMail's explicit modeling of social factors—such as relationship type, power dynamics, and cultural context—offers a potential mechanism to mitigate these biases by enabling users to specify culturally appropriate values. For instance, Western and Eastern students may configure different settings when communicating with their teachers \cite{den2005teacher}, thereby guiding the LLM toward culturally congruent tone generation. Future work should evaluate PersonaMail's effectiveness across diverse cultural populations and explore additional cross-cultural adaptation mechanisms, such as incorporating an AI-mediated "receiver perspective" feature that simulates how a recipient from a different cultural background might interpret the drafted message, helping users anticipate and adjust for potential cultural misalignments before sending.

\section{Conclusion}

The shift from generic LLM tools to systems like PersonaMail reflects a critical evolution in human-AI interaction: from efficiency-focused text generation to collaboration that honors the complexity of human communication. Our work suggests four actionable implications for future system design: (1) prioritize social context over generic fluency by modeling communication as situationally embedded practice; (2) align control with cognitive structures through intentional units (e.g., communicative units) rather than surface text; (3) treat user history as identity, not just data, learning stylistic patterns to preserve uniqueness; and (4) scaffold metacognition, not just production, helping users clarify social goals rather than outsource judgment.

PersonaMail demonstrates that AI can amplify human social intelligence by providing structure to tacit knowledge, granular control over expression, and continuity across interactions. As AI becomes more embedded in communication, the tools that succeed will be those that recognize a simple truth: In writing, as in life, it is not just what we say—but how we say it, to whom, and why—that matters.


\section{GenAI Usage Disclosure}
Generative AI is used in this paper for grammar suggestions. We have reviewed and edited the content as needed and take full responsibility for the content of this paper.

\begin{acks}
This work was (partially) supported by the Start-up Grant from City University of Hong Kong (Project No. 9610677).
\end{acks}

\bibliographystyle{acm-format/ACM-Reference-Format}
\bibliography{sample/sample-base}

\appendix
\section{Representative Systems}
\subsection{Representative Systems: Baseline Workflow} 

\subsubsection{MailMaestro}

MailMaestro (Figure~\ref{fig:formativestudy}-A) is a commercial email composition tool that provides AI-powered writing assistance through predefined tone categories (e.g., Formal, Professional) and adjustable length settings. Its key feature is a variable-based template system that allows users to save reusable email structures with substitutable fields for content such as names and dates, representing current commercial approaches to tone selection and adaptation.

\subsubsection{Friday Email} 

Friday Email (Figure~\ref{fig:formativestudy}-B) offers AI-assisted writing with 15 predefined tone options (e.g., formal, friendly, persuasive) and built-in editing tools—Extend, Rewrite, and Shorten—that support iterative refinement. It characterizes the commercial pattern of categorical tone control paired with multi-round AI revision features.

\subsubsection{Textoshop} 

Textoshop~\cite{masson_textoshop_2025} (Figure~\ref{fig:formativestudy}-C) draws inspiration from visual design software to enable direct manipulation of text tone. Its “Tone Picker” feature allows users to blend multiple tones and adjust intensity through continuous scales between paired opposites, offering fine-grained, composable tone control beyond discrete category selection.

\subsubsection{Gemini with Gmail (Baseline Workflow)} 

\begin{figure}[h]
  \centering
  \includegraphics[width=\linewidth]{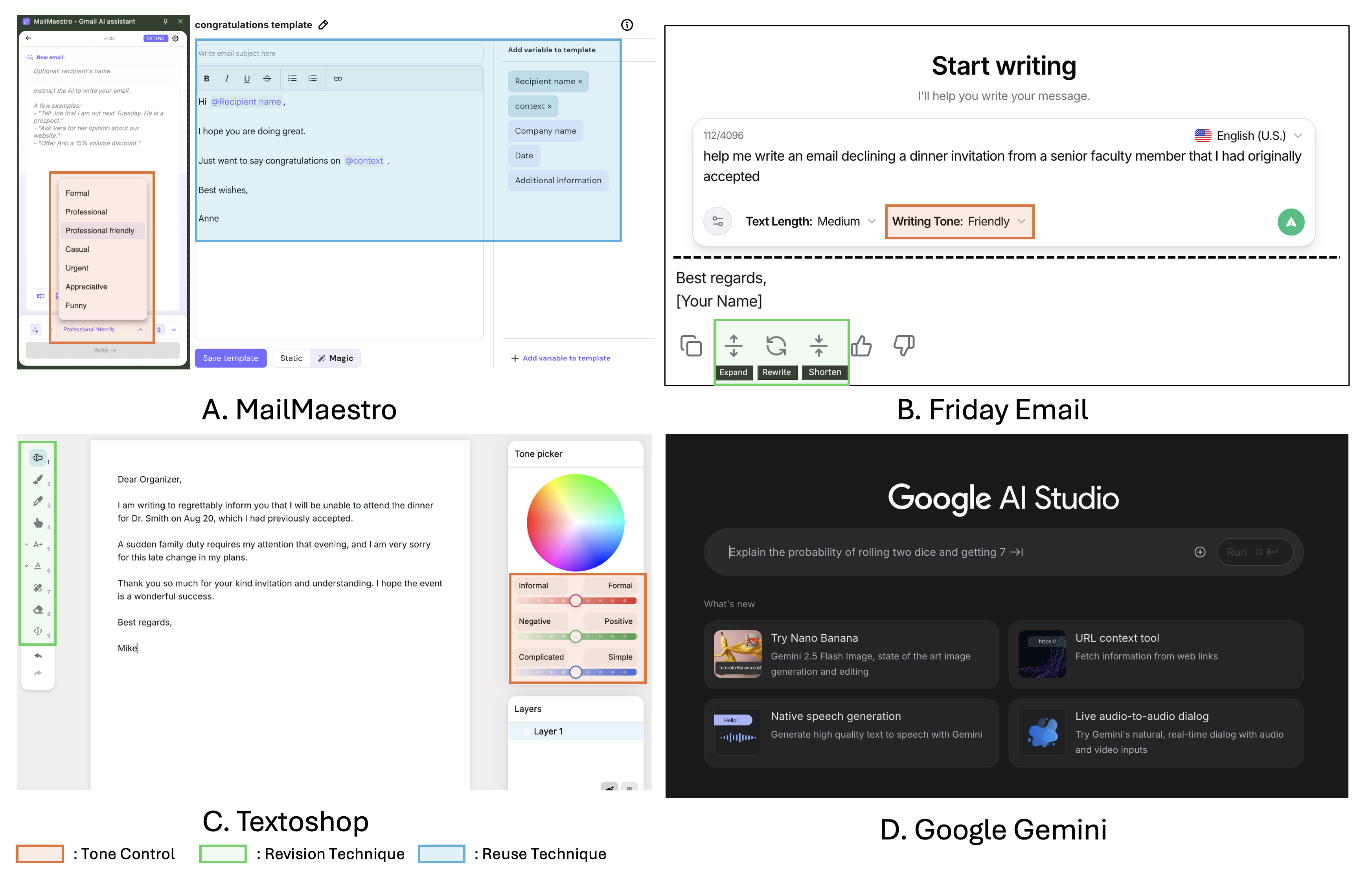}
  \caption{Representative systems used in our formative study spanning different approaches to LLM-assisted email composition.}
  \Description{A composite figure showing the user interfaces of different email composition systems used in the formative study, including MailMaestro, Friday Email, Textoshop, and Google Gemini paired with Gmail. }
  \label{fig:formativestudy}
\end{figure}

Gemini paired with the Gmail Editing Interface (Figure~\ref{fig:formativestudy}-D) represents the most common real-world workflow for AI-assisted email composition: conversational prompting followed by manual editing in a traditional email interface. We included it as a baseline to reflect everyday LLM use in its naturalistic form.

\vspace{3mm}

Together, these four systems capture the notable approaches encompassing LLM-assisted email composition: categorical tone selection, iterative revision, continuous tone blending, and conversational prompting. This range allowed us to examine how different design mechanisms influence user cognition during tone-sensitive writing.

\section{Classification of Existing LLM-Supported Email Writing and Tone-Setting Tools}
We classified existing LLM-supported email writing and tone-setting tools to understand the landscape of current approaches and identify design gaps that PersonaMail addresses.
Table~\ref{tab:designspace} presents this classification across three key dimensions: tone control mechanisms, revision techniques, and adaptation/reusability features.
\begin{table*}[t]
\small 
  \caption{Comparison of existing LLM-supported email tone-setting tools.}
  \label{tab:designspace}
  \begin{tabularx}{\textwidth}{p{1.9cm} p{3cm} p{4.5cm} p{4.1cm}}
    \toprule
    & \textbf{Tone Control} & \textbf{Revision Technique} & \textbf{Adaptation/Reusable Templates} \\
    \midrule
    MailMaestro\cite{nMailMaestro} & Predefined tone list & Manual edit, AI-based proofread & Text shortcut (e.g., mobile phone or address), variable-based template \\
    Friday Email\cite{FridayEmail} & Predefined tone list & AI tools (Extend, Rewrite, Shorten), multiple rounds of prompting & NA \\
    WriteMail\cite{WriteMail}  & Predefined tone list & NA & NA \\
    WordTune\cite{Wordtune} & Predefined tone list & Manual edit, AI tools (Variants, Extend, Shorten, Synonyms, Prompt) & NA \\
    Textoshop~\cite{masson_textoshop_2025} & Binary toggles, in-degree adjustment, combine tones & Manual edit, creative techniques inspired by drawing software & NA \\
    CDLR~\cite{zindulka_content-driven_2025} & Predefined tone categories (Positive, Neutral, Negative) & Manual edit, prompt-based refinement, AI improvement & NA \\
    \bottomrule
  \end{tabularx}
\end{table*}

\section{Systematic Review of Factors Influencing Written Communication Delivery}

Following the PRISMA guideline~\cite{page2021prisma}, we conducted a four-stage systematic review across three main databases: Web of Science, PsycInfo, and Scopus to identify empirical studies examining factors shaping written communication delivery in digital contexts (e.g., email, texting, messaging). This process led to the analysis of 97 qualified papers after manual screening from 1,946 related papers. The detailed process can be found in the Appendix B.

\paragraph{Search Strategy:}
We searched the databases using the query ((email OR "computer mediated communication" OR texting OR messaging) AND (tone OR style* OR nuance*) AND (factor OR reason* OR determinat* OR cause* OR perception OR context)). 
This structure intentionally balanced breadth (to include multiple CMC modalities) with specificity (to focus on factors influencing tone/style rather than general discourse).
The search was conducted in August 2025 and applied to peer-reviewed articles written in English, with no date restriction to ensure comprehensive coverage.
A total of 2,424 records were retrieved. 
After removing duplicates, 1,946 unique papers remained for screening.

\paragraph{Screening:}
Two researchers independently screened titles and abstracts. Studies were included if they empirically examined how written message delivery (e.g., tone framing, politeness, emoji use, sentence phrasing, or emotional expression) is shaped by individual, relational, or situational factors in everyday or interpersonal contexts.
We excluded studies at screening if they met any of the following:
(1) Non-empirical formats. Conceptual/theoretical essays, handbooks, viewpoints, and reviews without analyzable data (e.g., \cite{dada2022strategies}).
(2) Channel- or platform-centric work that examined adoption, usage, or feasibility of a medium (e.g., Facebook/WhatsApp uptake) without analyzing within-message delivery (e.g., \cite{lee2014mobile}).
(3) Macro/meso communication focused on campaigns, public messaging, or societal discourse (e.g., marketing, health promotion, political messaging, LGBTQ-inclusive communication, cybercrime/phishing detection) rather than interpersonal delivery features (e.g., \cite{jaworska2024making}).
(4) Technical or modeling papers (e.g., NLP/ML systems, detection/classification, conceptual architectures) that did not empirically examine human communication-style features or their perception (e.g., \cite{arfaoui2025unveiling}). 
(5) Verbal/prosodic or face-to-face cues (speech, facial expression) without a textual CMC component (e.g., \cite{jacob2012cerebral}).
This stage yielded 115 papers for full-text assessment.

\paragraph{Eligibility:}
In the eligibility scoping stage, we conducted a full-text assessment, where one researcher independently evaluated all 115 corpora, removing papers that did not meet the criteria and highlighting those requiring further discussion; a second reviewer resolved discrepancies. 
Additional full-text exclusions included: (1) Contact patterns rather than delivery: studies on when, how often, or who initiates messages without analysis of how messages are crafted (e.g., \cite{li2025correlates}); and (2) those overly fine-grained linguistic forms (phonological, syntactic, narrowly rhetorical, or lexical inventory counts) that did not link features to pragmatic/affective delivery (e.g., \cite{jacobs2023chimpanzee}); 
Following adjudication, 97 papers were retained for synthesis.

\paragraph{Synthesis:}
Through inductive coding and thematic clustering, we identified two overarching categories of communication-delivery determinants — Persona Factors and Situation Factors — encompassing 14 recurrent subdimensions in Table~\ref{tab:revised_factors_tone} (Note: When a factor is discussed in multiple works, only representative citations are included to conserve space). 

\begin{enumerate}[nosep]
    \item \textbf{Persona Factors} — attributes tied to the writer–recipient relationship, including relationship type, familiarity, power/status, gender dynamics, personality traits, relationship needs, age and culture.
      
    \item \textbf{Situation Factors} — contextual variables such as emotional intent, competing goals, promptness, communication purpose, occasion, and avoiding negative consequences.
\end{enumerate}

These categories together capture how writers interpret social context and interpersonal dynamics when shaping tone in digital correspondence. This taxonomy echoes Brown and Levinson's Politeness Theory~\cite{brown1987politeness}: our Persona Factors (Power/Status, Familiarity) operationalize their dimensions of Power ($P$) and Social Distance ($D$), while Situation Factors (Communication Purpose, Emotional Intent) capture Ranking of Imposition ($R$), grounding PersonaMail's design in established face-work principles.

\section{Quantitative Metrics of Evaluation}
\textbf{Email Quality and Articulation Metrics (RQ1)}: To evaluate whether PersonaMail's factor-based scaffolding enables more comprehensive articulation and higher-quality outputs, we collected both quality ratings and articulation depth metrics. Participants rated email quality using a 7-point Likert scale (1 = Least Satisfied, 7 = Most Satisfied) at two critical stages: immediately after initial AI generation (first draft quality) and after completing their revisions (revised draft quality). For PersonaMail, we assessed articulation comprehensiveness by tracking the number of communication factors engaged with during factor exploration, the total word count of prompts and factor responses, and the time spent on prompting and factor exploration (articulation time before AI generation). For the baseline system, we collected comparable proxy metrics: the number of communication aspects mentioned in free-form prompts (manually coded from prompt content), total prompt word count, and prompting time, enabling direct comparison between PersonaMail's structured approach and the baseline's open-ended prompting.

\textbf{Control and Revision Efficiency Metrics (RQ2)}: To evaluate how effectively PersonaMail's multi-level control architecture supports efficient and precise revision, we collected detailed temporal and behavioral interaction data throughout the composition process. We measured revision time (time spent modifying AI-generated drafts to reach satisfactory quality) for both systems. For PersonaMail, we logged the distribution of revision operations across three categories: intent-driven modifications, Quick Fix applications, and traditional text editing, and calculated acceptance rates for system suggestions by tracking the percentage of intent modifications and Quick Fix suggestions that users accepted versus ignored. For the baseline system, we tracked revision operations categorized as conversational AI prompting for modifications and manual text editing in Gmail, providing comparable data on revision strategies and control mechanisms.

\textbf{Adaptive Learning and Reusability Metrics (RQ3)}: To assess whether PersonaMail's adaptive features—Persona-Situation Anchors and the Adaptive Stylebook—reduce articulation burden and improve efficiency without compromising quality, we tracked changes across PersonaMail's three usage rounds (Rounds 1-3). Specifically, we measured the number of factors engaged with in each composition session and the number of factors requiring revision when using Persona-Situation Anchors. We also tracked temporal metrics (prompting time, revision time, total task time) across rounds to quantify efficiency gains. Quality ratings across the three rounds were compared to ensure adaptive features maintained output quality standards.

\textbf{Cognitive Load Metrics (RQ4)}: To evaluate whether PersonaMail's structured approach reduces cognitive burden compared to open-ended conversational AI interfaces, we measured subjective cognitive load using the NASA Task Load Index (TLX) \cite{hart1988development} after each email composition task. Participants answered the six TLX items using a 10-point Likert scale. We calculated the Raw-TLX score \cite{hertzum2021reference} as the simple average of the six scales, where higher scores indicate greater cognitive load. Participants completed the TLX immediately following task completion for both baseline and PersonaMail conditions in Round 1, enabling direct comparison of cognitive load between the two approaches.

\section{Writing Tasks for Formative Study}

\subsection{Workplace Writing Tasks}

\begin{enumerate}
    \item \textbf{Salary Negotiation with HR:} You are a recent graduate writing to the HR department of a company to renegotiate your base salary for a job offer. It’s your only offer, and you have a good relationship with the recruiter, so you must persuade HR to raise the salary without risking the withdrawal of the offer.
    
    \item \textbf{Mid-Level Employee Pay Raise Request:} As a mid-level employee during a company pay freeze, write to your manager to request a raise or a clear promotion plan. You may reference another job opportunity as leverage, but without sounding like you’re issuing an ultimatum.
    
    \item \textbf{Project Leader Deadline Enforcement:} As a project leader, write to a collaborator who is also a friend to firmly request overdue key data submission before an important deadline—without harming your friendship.
    
    \item \textbf{Correcting a Published Paper:} As a paper author, formally inform the journal editor that your published paper contains a serious error. Ask about the correction or retraction process while addressing your coauthor’s concerns about reputation.
    
    \item \textbf{Addressing a Colleague’s Negativity:} As a team member, privately communicate with a coworker and friend whose negative attitude has affected team morale. Your goal is to address the issue honestly while preserving the friendship.
    
    \item \textbf{Handling Employee Lateness:} As a direct supervisor, email an employee who has recently developed a pattern of tardiness. They were previously punctual and high-performing. Address the issue clearly but without being overly harsh.
    
    \item \textbf{Work-Life Balance Advice:} Write to a hardworking collaborator who appears stressed and overworked. You want to show concern and offer support, even though the project deadline remains tight.
    
    \item \textbf{Mediating Team Conflict:} Two team members who usually cooperate well have recently argued publicly during meetings. Write to mediate the situation and restore a professional and comfortable team environment.
    \item \textbf{Declining a Professional Dinner Invitation:} Politely decline an important career-related dinner invitation due to unavoidable personal matters, while minimizing disappointment to the inviter whom you value.
\end{enumerate}

\subsection{Educational Writing Tasks}

\begin{enumerate}
    \item \textbf{Scholarship Extension Request:} As a PhD student facing a family emergency, write to your advisor to request a three-month extension of your scholarship, which is about to expire. Your parent is ill, and your advisor has previously expressed impatience with your research progress.
    
    \item \textbf{Authorship Order Change Request:} As a PhD student, request that your advisor reconsider the authorship order on a paper where you feel your contribution was undervalued.
    
    \item \textbf{Reporting TA Misconduct:} As a graduate student, write to your university’s HR or relevant office to report ongoing microaggressions or harassment by a teaching assistant. Although evidence is limited, you request confidentiality and temporary protective measures.
\end{enumerate}

\subsection{Family and Personal Writing Tasks}

\begin{enumerate}

    \item \textbf{Declining Free Professional Work for a Friend:} As a professional graphic designer, politely decline a close friend's request for free design work on their startup idea. They promise future equity, but the project is high-risk and you're already overworked. Refuse tactfully without dampening their enthusiasm.
    
    \item \textbf{Setting Financial Boundaries with Family:} Write an email to a close sibling who requests a large loan for a risky business venture. They have a history of poor financial management and unpaid small loans. You must refuse the request kindly but firmly.
    
    \item \textbf{Roommate Cleanliness Discussion:} Have an open and respectful conversation with your roommate (not a family member) about hygiene and cleanliness issues that have been affecting your shared living space.
    
    \item \textbf{Discussing Future Differences with a Partner:} You and your long-term partner are about to graduate but have conflicting plans regarding location, career, or further study. Write a dialogue or reflection exploring how to discuss and manage these differences.
\end{enumerate}

\section{Writing Tasks for Adaptive Reuse Evaluation and Task Validation Process}
\subsection{Representative Tasks}
\label{appendix:representative_tasks}
The following five task groups demonstrate the tasks used in our PersonaMail repeated use study:

\textbf{Task Group 1: Addressing Personal Boundary Violations in Shared Spaces.} This task group focused on confronting uncomfortable behaviors in close-proximity living or working environments:
\begin{itemize}
  \item Addressing a roommate whose poor hygiene habits (not cleaning shared areas, accumulating personal items, creating unpleasant odors) are severely impacting living quality
  \item Speaking with an office neighbor whose disruptive workspace habits (loud mechanical keyboard typing, music leaking from headphones, eating strongly scented food at their desk) are interfering with concentration
  \item Requesting a nearby female colleague to reduce her use of a strong, inexpensive perfume that causes headaches and dizziness during close interactions
\end{itemize}

\textbf{Task Group 2: Negotiating Recognition and Fairness in Academic Settings.} This task group addressed situations where contributions were undervalued or misrepresented in contexts with power imbalances:
\begin{itemize}
  \item A postgraduate student requesting their advisor to reconsider author order on a publication where their contributions were underestimated
  \item A student asking a professor to reassess individual contributions and grading in a group project where they carried the primary workload but received scores that undervalued their effort
  \item An intern requesting their supervisor to strengthen or revise a recommendation letter that was overly generic and failed to highlight specific accomplishments and contributions
\end{itemize}

\textbf{Task Group 3: Declining Requests from Close Relationships.} This task group focused on setting boundaries with friends and family without damaging important relationships:
\begin{itemize}
  \item A professional graphic designer declining a close friend's request for free design services for a high-risk entrepreneurial venture, despite promises of future equity or compensation
  \item Refusing a sibling's request for a substantial loan for a risky business investment, given their history of financial instability and unpaid previous debts
  \item Declining a friend's urgent request for an internal job referral when the requester's professional capabilities and work habits are misaligned with the company's culture and standards
\end{itemize}

\textbf{Task Group 4: Negotiating Employment Terms with Limited Leverage.} This task group involves situations where individuals must advocate for better terms while maintaining goodwill and avoiding jeopardizing opportunities:
\begin{itemize}
  \item A recent graduate writing to the HR department of a company to renegotiate the base salary for a job offer. This is the only offer received, and the relationship with the recruiter is positive, creating concern that aggressive negotiation might feel like "asking a friend for money." The task is to persuade HR to increase the salary while avoiding withdrawal of the offer.
  \item A recent graduate requesting an extension of the decision deadline for a job offer from a top company. The company requires a response within 5 days, but the candidate is awaiting final interview results from a more preferred company in 7 days. This first offer serves as a safety net and cannot be lost. The email must politely request a two-day extension from the friendly HR recruiter without making them feel like a "backup option" and risking offer withdrawal.
  \item A current employee requesting adjustment to work hours or transition to remote work due to personal reasons (such as family needs or health issues). While the relationship with the supervisor is friendly, company policy is typically strict. The email must persuade management or HR to approve the adjustment while avoiding being perceived as uncommitted or negatively impacting career prospects.
\end{itemize}

\textbf{Task Group 5: Enforcing Commitments in Friend-Based Collaborations.} This task group addresses situations where personal relationships complicate the enforcement of professional or practical obligations:
\begin{itemize}
  \item As a project leader, writing to a collaborator who is also a friend to request overdue submission of critical data before an important deadline, while maintaining the friendship.
  \item Writing to a roommate who is also a close friend to request completion of their assigned cleaning responsibilities in shared living areas.
  \item Writing to your best friend, to whom you rented your apartment, who has now missed two months of rent payments and consistently makes excuses when contacted. The email must inform them that rent must be paid in full by a specified date.
\end{itemize}
\subsection{Task Validation}
\label{appendix:task-validation}

To ensure task quality, thematic coherence and difficulty consistency across adaptive-use conditions, we employed the following validation process.

The first author initially prepared sets of email writing tasks based on communication-related literature and personal experiences, which were then independently evaluated by two expert researchers. These experts rated each task set on two critical dimensions using 5-point Likert scales: (1) difficulty similarity across the three tasks within each set, and (2) contextual relevance—the degree to which tasks shared common relationship dynamics or situational challenges that would enable meaningful reuse of communication strategies. They also provided qualitative comments identifying specific reasons when difficulty or contextual relevance appeared inconsistent within a task set. Based on this feedback, the first author iteratively revised the writing tasks until both difficulty similarity and contextual relevance achieved ratings of 4 or above for all tasks within each set.

\end{document}